\documentclass[12pt]{article}

\usepackage{amsmath,amssymb,amsbsy,amsfonts,latexsym,graphicx}
\usepackage{hyperref}
\usepackage{color, array, subfigure}
\usepackage{cite}

\allowdisplaybreaks

\evensidemargin \oddsidemargin

\begin{document}


\begin{titlepage}

\renewcommand{\thefootnote}{\fnsymbol{footnote}}



\vspace{15mm}
\baselineskip 9mm
\begin{center}
  {\Large \bf 
{  
Rotating End of the World}
 }
\end{center}

\baselineskip 6mm
\vspace{10mm}
\begin{center}
Kyung Kiu Kim$^{a,d}$, Hwajin Eom$^{b,d}$, Jung Hun Lee$^{c,d}$ and Yunseok Seo$^{a,d}$
 \\[10mm] 
  {\sl ${}^a$College of General Education, Kookmin University, Seoul 02707, Korea}
   \\[3mm]
  {\sl ${}^b$Dasan University College, Ajou University, Suwon-si, Gyunggi-do 16499, Korea}
   \\[3mm]
  {\sl ${}^c$Department of Physics, Jeonbuk National University, Jeonju, 54896, Korea}
   \\[3mm]
  {\tt kimkyungkiu@kookmin.ac.kr, eom16@ajou.ac.kr, Junghun.lee@jbnu.ac.kr, yseo@kookmin.ac.kr
  }
\end{center}
\thispagestyle{empty}

\vspace{1cm}
\begin{center}
{\bf Abstract}
\end{center}
\noindent
We study the thermodynamics and interior structures of dynamical end of the world (EoW) branes in the rotating BTZ black hole. By mapping the induced metric of the branes to an effective Jackiw-Teitelboim (JT) system, we derive the first law of thermodynamics for the boundary conformal field theory (BCFT), incorporating boundary degrees of freedom. To construct this thermodynamics, we adopt two frameworks of black hole chemistry and the duality between the JT black hole and the Sachdev-Ye-Kitaev (SYK) model. In addition, we verify that the shadow entropy is equivalent to the boundary entropy via Hubeny-Ryu-Takayanagi (HRT) surfaces. Furthermore, we explore the possible interior configurations of the EoW brane inside the horizon. Two representative configurations, namely single and double-joint EoW branes, are considered. These disparate configurations share the same exterior brane configuration outside the horizon. By comparing their energies, we show that a transition could occur within the horizon.
\\ [15mm]
Keywords: Gauge/gravity duality, BCFT, Thermodynamics, End of the world brane. 

\vspace{5mm}
\end{titlepage}

\baselineskip 6.6mm
\renewcommand{\thefootnote}{\arabic{footnote}}
\setcounter{footnote}{0}

\section{Introduction}

The holographic approach using gauge/gravity duality \cite{Maldacena:1997re, Aharony:1999ti, Maldacena:2011ut} has provided a profound framework for understanding an invisible link between gravity and quantum field theory. This framework has been extended to various areas of theoretical physics. In particular, this approach is useful for studying BCFT, the conformal field theory on a manifold with boundaries \cite{Cardy:1984bb, McAvity:1995zd}. The boundary condition is chosen to be consistent with the conformal symmetry \cite{Cardy:1989ir}. This boundary carries information and its entropy showing the g-theorem \cite{Affleck:1991tk, Friedan:2003yc}. The gravity dual of BCFT is a curved spacetime with boundaries, in which a special kind of brane is located. This brane is dubbed the EoW brane, which imposes a Neumann-like boundary condition, physically truncating a part of the full spacetime.

This holographic BCFT configuration using EoW branes was first suggested in \cite{Takayanagi:2011zk} as an extension of AdS/CFT correspondence. The corresponding gravity dual geometrically realizes the boundary entropy and the g-theorem with a tension parameter of an EoW brane \cite{Fujita:2011fp, Nozaki:2012qd}. Recently, the AdS/BCFT setup has played a pivotal role in resolving the black hole information paradox by providing a natural environment for the emergence of islands \cite{Almheiri:2019hni}\footnote{See \cite{Geng:2020qvw, Geng:2020fxl, Geng:2021iyq, Geng:2021wcq} for various generalizations of this framework, including the effects of graviton mass, gravitating baths, and communicating black hole systems.}. See \cite{Almheiri:2020cfm} for a recent review. Thus, it is legitimate to consider an EoW brane inside a black hole horizon. Also, the EoW brane is a fundamental building block for extending holography for AdS space to other spacetimes, e.g., \cite{Takayanagi:2011zk, Miyaji:2015yva, Cooper:2018cmb, Geng:2021wcq, Hao:2025ocu}. Microscopically, such EoW branes can be understood as representing specific pure states in the dual SYK model, providing a fundamental bridge between boundary states and bulk interiors \cite{Kourkoulou:2017zaj}.

For the reasons mentioned, we regard the EoW brane as an important probe for studying black hole physics, such as thermodynamics and the black hole interior. Since the original BCFT studies \cite{McAvity:1995zd, Behrend:1999bn, Friedan:2003yc, Cardy:2004hm} consider two-dimensional systems with a temperature circle, it is natural to consider a three-dimensional Euclidean or Lorentzian black hole as a gravity dual. In this work, we will focus on the Lorentzian black hole. In our previous study \cite{Kim:2023adq}, we studied the thermodynamics of the BCFT system using the thermodynamics of the JT black hole and the non-rotating BTZ black hole, and we investigated the EoW brane dynamics inside the horizon\footnote{The rotating BTZ metric becomes subtle when we take the Euclidean version. Some components of the metric are not real. This is one of the reasons why we concentrate on the Lorentzian black hole.}. We now extend and develop the previous result to the case of the rotating BTZ black hole.

The rotating BTZ black hole is a crucial object in the AdS/CFT. This is dual to the 2D CFT with a nonvanishing momentum flow $T^{t\phi}$ along a periodic spatial direction. This flow makes the right and left movers of the CFT distinct, so they feel like formally different temperatures. If we locate two boundaries along this spatial direction, we cannot avoid a source term and a sink term at these two boundaries to maintain the stationary flow. Then, one may naturally ask what kind of gravity dual can explain these sink and source operators, and how one can build a thermodynamics for this system. We will identify the corresponding gravity duals with two moving EoW branes in the rotating black hole. As we mentioned, an EoW brane effectively truncates a part of the spacetime, thereby providing a geometric cutoff. As these branes propagate through the BTZ bulk, they effectively dynamically reconfigure the spacetime manifold by `paving' or `annihilating' regions of the geometry. This process provides a classical geometric manifestation of the underlying quantum gravity dynamics inherent in the BCFT with nonvanishing momentum flow.

Also, the obtained EoW brane configuration can be interpreted from the BCFT perspective. A basic understanding of such a system can be analyzed using thermodynamics. Thus, we construct the first law for this stationary system. One of the important ingredients in the first law is how to encode the boundary degrees of freedom into the thermodynamic relation. The boundaries of the BCFT are also the boundaries of two JT black holes. Therefore, one can apply the second holography to these boundary points.  We adopt the thermodynamics of the JT black hole to describe the boundary state of the BCFT. Since this JT system has a tension-dependent cosmological constant, one may use the so-called black hole chemistry \cite{Kastor:2009wy, Dolan:2010ha, Dolan:2011xt, Cvetic:2010jb, Kubiznak:2012wp, Kubiznak:2016qmn}. This idea was used in our previous work \cite{Kim:2023adq}. In the present work, we reformulate the thermodynamics in standard form by defining a new thermodynamic volume. We also propose another expression of the first law based on the duality between the SYK model and the JT black hole \cite{Sachdev:1992fk, Kitaev:2015, Maldacena:2016hyu, Sarosi:2017ykf, Berkooz:2018jqr, Lin:2022rbf}. In this first law, an average strength $\mathcal{J}_{\text{SYK}}$ of the random coupling is the most important variable. Therefore, we apply the second holography to the boundary of BCFT as an effect of the SYK model.

In our thermodynamic analysis, the shadow entropy plays an important role. This entropy equals the area of the region cast on the horizon obscured by the EoW branes. It was known that the boundary entropy is determined through the Island/BCFT correspondence \cite{Suzuki:2022xwv}. In the non-rotating BTZ black hole, the boundary entropy is the area of a particular minimal surface. Also, this area is the same as the shadow entropy. See Appendix in \cite{Kim:2023adq} for an explicit calculation. We extend this equivalence to the case of the rotating black hole. We find relevant HRT surfaces consistent with the boundary entropy in  \cite{Suzuki:2022xwv}  and the shadow entropy. Therefore, we regard the total horizon area, including shadows of the EoW branes, as the total entropy of the BCFT system.

In more detail, the EoW branes border the truncated region and the spacetime dual to BCFT. Furthermore, this brane touches the black hole horizon smoothly. This implies the EoW brane can probe the black interior if it doesn't touch the bulk singularity. Also, an EoW brane may contain a quantum extremal surface and an island that are commonly placed inside a black horizon.   In our previous work \cite{Kim:2023adq}, we tried to reach a physical implication using two EoW branes. By then, we used a simple scale argument with the location of a touching point by two EoW branes inside the horizon.

As a new ingredient for the present work, we introduce a nonsmooth configuration, a cusp of an EoW brane. This is called a joint, characterized by a highly localized energy density on the EoW brane \cite{Hayward:1993my}. This nonsmooth configuration is nothing but a junction of a junction (EoW brane). See \cite{Biswas:2022xfw} for a more extensive application. To try an advanced analysis of the EoW brane, we regard two EoW branes touching a point as a single brane with a joint. We study the features of this joint sitting on an EoW brane inside the horizon.

Using this building block, we find the simplest and most representative interior configurations. Two types are considered. One is the single-joint configuration, and the other is the double-joint configuration with a spacelike EoW brane. We show that an outer product of adjacent normal vectors gives the joint energy. Also, we compare the energies of two configurations to provide a clue to a possible transition inside the horizon.

The organization of this paper is as follows. In Section 2, we review the thermodynamics of the rotating BTZ black hole and describe the dynamics of rotating EoW branes within this background. In Section 3, we derive the first law of thermodynamics for the BCFT system with two boundaries. To do this, we take two strategies. One is to use the ``thermodynamic volume''; the other is to use the duality between JT gravity and the SYK model. We also verify the equivalence between the shadow and boundary entropy using HRT surfaces. In Section 4, we explore the interior configurations of the EoW branes, including single and double-joint configurations. Finally, we conclude with a discussion of the implications for the boundary theory in Section 5.

\section{End of the World in Rotating BTZ Black Hole}

This section summarizes the thermodynamics of the BTZ black hole. In addition, we construct a BCFT system with a nonvanishing momentum flow by combining a pair of EoW branes that move in the bulk. An EoW brane solution, as found in \cite{Fujita:2011fp}, serves as a basic building block for our main construction.

\subsection{Thermodynamics of the rotating BTZ black hole}\label{sec: BTZ thermodynamics}

Let us consider the holographic BCFT associated with the rotating BTZ black hole. The corresponding gravity dual is given by a truncated spacetime with EoW branes on its boundaries. We start with the well-known metric as follows:
\begin{equation}
ds^2  = -\mathcal{U}(r)\, dt^2 +\frac{dr^2}{\mathcal{U}(r)} +r^2 \left(d\phi-\frac{JL}{2r^2}dt\right)^2~~ \text{with}\quad \mathcal{U}(r)=\frac{r^2}{L^2}-M+\frac{J^2L^2}{4r^2}\,.
\end{equation}
The inner and outer horizons can be found in terms of the mass and angular momentum as
\begin{equation}
r_\pm^2 =\frac{M L^2}{2} \left(1\pm \sqrt{1-\frac{J^2}{M^2}}\right)\,.
\end{equation} 
The temperature and entropy are given by
\begin{align}\label{T S}
T =\frac{r_+^2-r_-^2}{2 \pi  L^2 r_+} ~~,~~S = \frac{\mathcal{A}_{2\pi}}{4G}=2\pi L s = \frac{2\pi L r_+}{4G L}\,,
\end{align}
where $s$ is the entropy density and $\mathcal{A}_{2\pi}$ denotes the horizon area associated with the full angle $2\pi$. This superfluous notation is due to the later use of the BCFT volume associated with an angle $\Delta\phi$.

Since we denote the spatial volume of the BCFT by $\mathcal{V}=L\Delta\phi $ and we will vary the volume, we take a fan-shaped part of the BTZ black hole with angle $\Delta\phi$ for later convenience. Also, densities are more useful for this reason. In particular, the energy and the angular momentum densities can be evaluated from the holographic energy-momentum tensor. They are $\epsilon\equiv \left<T_{tt}\right>$ and $j \equiv -\left<T_{t\phi}\right>$, respectively. See Appendix A for detailed derivations. In addition, the angular velocity is given by $\Omega = \frac{1}{L}\frac{r_-}{r_+}=\omega/L$. Using these quantities, one can find the first law for the fan-shaped boundary system as follows:
\begin{align}\label{first law without boundary}
\delta\mathcal{E}_{\mathcal{V}} = T \delta S_{\mathcal{V}} +\Omega \delta \mathcal{J}_{\mathcal{V}} -  \mathcal{P}\delta\mathcal{V}\,,
\end{align}
where the extensive quantities such as energy, entropy, and angular momentum can be written as
\begin{align}
\mathcal{E}_{\mathcal{V}} =  \epsilon\mathcal{V}~,~S_{\mathcal{V}}=\frac{\mathcal{A}_{\Delta\phi}}{4G}= s \mathcal{V}~,~\mathcal{J}_{\mathcal{V}}=j \mathcal{V}\,.
\end{align}
The pressure $\mathcal{P}$ is given by the on-shell action, as shown in Appendix A.

\subsection{A pair of rotating EoWs: creating and annihilating spacetime}

This section reviews EoW brane solutions on the rotating BTZ black hole. A single solution was found in \cite{Fujita:2011fp}. We introduce another familiar coordinate, $ z=L/r$, just for convenience from this section on. Then, the metric is written as
\begin{align}\label{metric BTZ}
ds^2 =  \frac{L^2}{z^2}
\left[-f(z)d\tilde{t}^2 +\frac{dz^2}{f(z)} +\left(d\phi -\frac{z^2 }{z_+ z_-}d\tilde{t} \right)^2\right],
\end{align} 
where the metric function is 
\begin{align}
f(z)= \left(1-\frac{z^2}{z_+^2} \right)
\left(1-\frac{z^2}{z_-^2} \right)\,.
\end{align}
Here, $z_{-}\geq z_{+}$ and the dimensionless time coordinate is defined as $\tilde{t}=t/L$. The parameters $M$ and $J$ for the mass and the angular momentum are given in terms of the inner and outer horizons as follows:
\begin{align}
M=\frac{1}{z_+^2} +\frac{1}{z_-^2}~~,~~ J=\frac{2}{z_+ z_-}\,.
\end{align}

In general, the location of the EoW brane can be taken as $\phi = g(\tilde{t},z)$. Thus, it is desirable to introduce a coordinate given by $\psi\equiv g(\tilde{t},z) -\phi$. Along this coordinate, one can decompose the metric (\ref{metric BTZ}) as follows:
\begin{align}
ds^2 = N^2 d\psi^2  +  h_{ab}\left( dx^a + V^a d\psi \right)\left( dx^b + V^b d\psi \right)\,,
\end{align}
where $x^a=(\tilde{t}, z)$ and $a=1,2$. The induced metric, shift vector, and lapse function can be read from the above as
\begin{align}\label{ind_met shift lapse}
&h_{ab} = L^2 \left(
\begin{array}{cc}
 \frac{\left(z^2 J-2 \dot{g}\right)^2-4 f}{4 z^2} & -\frac{g' \left(z^2 J-2 \dot{g}\right)}{2 z^2} \\
 -\frac{g' \left(z^2 J-2 \dot{g}\right)}{2 z^2} & \frac{1+\left(g'\right)^2 f }{f z^2} \\
\end{array}
\right),\,V_a= L^2 \left( \frac{J}{2}-\frac{\dot{g}}{z^2},\, -\frac{g'}{z^2} \right)\nonumber\\
&N^2 =\frac{4 L^2 f }{z^2 \left[4\left(g'\right)^2 f^2 +4 f-\left(z^2 J-2 \dot{g}\right)^2\right]}\,,
\end{align}
where $\dot{}$ and ${}'$ denote the derivative with respect to $\tilde{t}$ and $z$, respectively. The EoW brane at $\psi=0$, with a constant tension $\sigma/L$, obeys the following junction equation:
\begin{align}\label{junction eq 0}
K h_{ab}-K_{ab} = T_{ab}= -\frac{\sigma}{L}  h_{ab}\,.
\end{align}
Also, one can decompose the equation into its trace and traceless parts. So, we demand  $2K_{ab}-K h_{ab}$ vanish first, and then, we may solve the trace part $K = -2 \sigma/L$.This junction formulation is based on the probe limit of the branes; therefore, we consider the parameter range $0<\sigma<1$. A more detailed discussion of the parameter range or probe limit is provided in Appendix B.

\begin{figure}
        \centering
        \begin{subfigure}{}
            \centering
            \includegraphics[width=150mm]{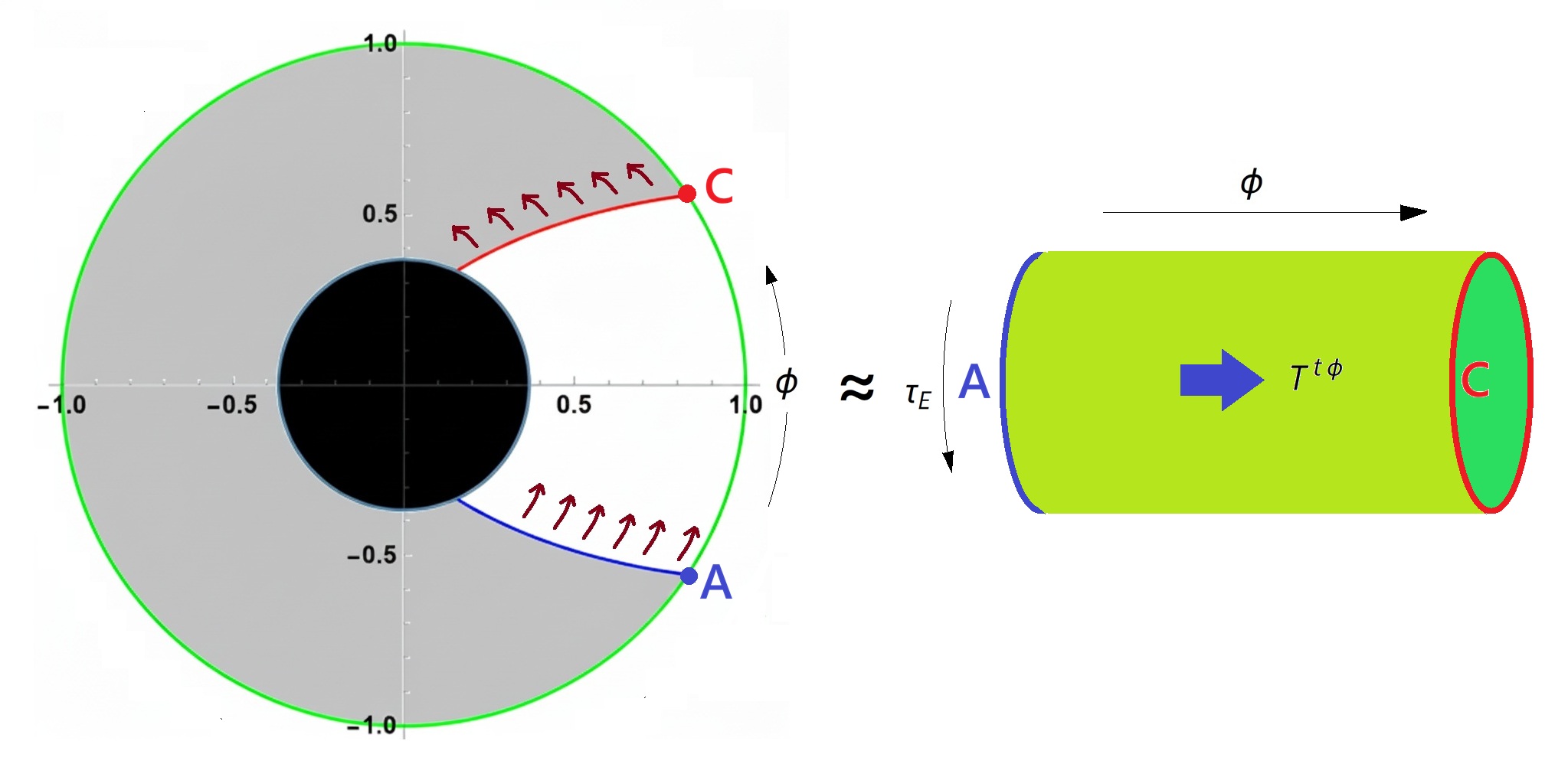}
        \end{subfigure}
                 \caption{{\bf Rotating EoW branes(Left) and CFT dual(Right):} The CFT dual carries a momentum flux $T^{t\phi}$ along the angle coordinate. To make the system stationary, it should source and sink the flux at A and C, respectively. The left figure shows the EoW branes touch A and C simultaneously, thereby truncating the gray region. By the rigid rotation, the red and blue branes are creating (paving) and annihilating (removing) spacetime, respectively. }\label{fig: EoW cartoon00}
\end{figure}

As we mentioned earlier, a relevant solution for this junction equation was provided in \cite{Fujita:2011fp}. This solution takes the form as $g(\tilde{t},z)= \mathbb{C}_0 \tilde{t} + \Phi(z)$. Here, $\mathbb{C}_0$ can be $z_-/z_+$ or $z_+/z_-$. Since the former is the superluminal EoW case, we choose the latter. Thus, the junction profile $\phi - (z_+/z_-)\, \tilde{t}=\Phi(z)$ implies a rigidly rotating brane. Then, the traceless part of the junction equation (\ref{junction eq 0}) is reduced to the following differential equation:
\begin{align}
\Phi''(z)=\frac{3 z \Phi'(z)}{z_-^2-z^2}- \frac{z \left(z_-^2-z^2\right) \Phi'(z)^3}{z_-^2 z_+^2}\,.
\end{align}
This equation admits a solution given by
\begin{align}
\Phi'(z)= \pm \frac{z_- z_+}{\left(z_-^2-z^2\right) \sqrt{1+\mathbb{C}_1 \left(z_-^2-z^2\right) z_-^2 z_+^2}}\,,
\end{align}
where $\mathbb{C}_1$ is an integration constant. Plugging this into the trace part of the junction equation (\ref{junction eq 0}), the location of the EoW brane is determined as follows:
\begin{align}\label{dG sol}
\Phi'(z) = \pm \frac{  \sqrt{1-\frac{z_+^2}{z_-^2}}\,\sigma}{\left(1-\frac{z^2}{z_-^2}\right) \sqrt{\left(1-\frac{z^2}{z_-^2}\right)-\sigma^2 \left(1-\frac{z^2}{z_+^2}\right)}}\,.
\end{align}
Here, $\sigma$ ranges $0<\sigma<1$. We only consider a positive tension in the present paper for simplicity.  The signs of the above imply that one can find two possible EoW branes anchored at two different boundary positions, $\Phi(0)=\phi-z_+/z_-\tilde{t}$. Since the positive tension for each brane is supposed, the sign denotes the direction of the normal vector of each brane. Therefore, to have an inner region of two EoW branes as a gravity dual like the cartoon of Figure \ref{fig: EoW cartoon00}, a pair of solutions with opposite signs must be considered. We will denote these two solutions by $\Phi_\pm(z)$.

In general, each brane in the pair can have a different tension,
which we denote by $\sigma^i$ $(i=1,2)$.
The junction condition \eqref{junction eq 0} is then imposed independently on each brane, and its radial profile is obtained from \eqref{dG sol} by replacing $\sigma$ with $\sigma^i$. Within the class of solutions with $\mathbb{C}_0=z_+/z_-$ chosen above, both branes share the same rigid angular velocity, while their radial profiles generally differ for unequal tensions, $i.e.$, $\sigma^1\neq\sigma^2$. For simplicity of the configuration illustrated in Figure 1, we take the same tension \(\sigma\) for both branes in the following.
In addition, without loss of generality, we may fix the boundary points with $\tilde{t}=0$ as $\Phi_\pm(0) = \pm \Delta\phi/2$. On the other hand, the above solutions become $\Phi_\pm'(z)\to \pm\,\sigma/\sqrt{1 -\sigma^2 (1-z^2/z_+^2)}$ in the limit of the non-rotating BTZ black hole ($z_-\to\infty$). This is the same as the well-known solution \cite{Takayanagi:2011zk}. See \cite{Kim:2023adq} for the explicit form in our notation. Coming back to the rotating case, we write the locations of the EoW branes by further integration as
\begin{align}\label{EoW position}
\phi-\frac{z_+}{z_-}\tilde{t}=\Phi_{\pm}(z) =  \pm \frac{\Delta\phi }{2} \pm z_+ \tanh ^{-1}\left(\frac{ \sqrt{1-\frac{z_+^2}{z_-^2}}\,\sigma  }{ \sqrt{\left(1-\frac{z^2}{z_-^2}\right)-\sigma ^2 \left(1-\frac{z^2}{z_+^2}\right)}} \frac{z}{z_+}  \right)\,.
\end{align} 
For convenience and due to the periodicity of $\phi$, the locations of the two EoW branes are also chosen by
\begin{align}\label{EoW rotation}
\phi - \frac{z_+}{z_-} \,\tilde{t} = \Phi_+(z)~~,~~\phi - \frac{z_+}{z_-} \,\tilde{t} = 2\pi + \Phi_-(z)\,.
\end{align}

Let us look at Figure \ref{fig: EoW cartoon00}. When $J=0$, the dual boundary state is a thermal state without the momentum flow $\left<T^{t\phi}\right>$ along the $\phi$-direction, and the two EoW branes are at rest. Now, let us suppose that we turn on $J$ of the BTZ black hole. Then the momentum starts to flow from one boundary to the other boundary. This implies that the two boundaries begin to serve as a momentum generator and a momentum sink, respectively. Also, both EoW branes begin to rotate and sweep through the entire spacetime. However, the roles of these two branes are different. Under the rotation, the leading brane C in the figure is creating the spacetime region, whereas the following brane is annihilating the spacetime region under the rotation. Therefore, one may imagine that the following brane burns a spacetime region like a fuel, and this burning can produce the momentum flow in the dual field theory. On the other hand, the momentum flow is simultaneously eaten by the boundary C, and the leading brane is now creating a spacetime region at the cost of the momentum flow.

At an instant, each EoW brane anchors at a boundary point A or C at $z=0$ and touches the horizon. This brane profile casts a shadow that reaches the horizon and creates an additional area of the horizon. The additional angle covering the area on one side is $\Phi_+(z_+)-\frac{\Delta\phi}{2} = z_+ \tanh^{-1}\sigma$. Therefore, the corresponding shadow entropy is 
\begin{align}\label{shadow entropy}
S_{shd} = \frac{L}{4G} \tanh^{-1}\sigma\,.
\end{align}
Interestingly, this coincides with the result for the non-rotating BTZ case \cite{Kim:2023adq}. In the non-rotating case, the boundary entropy is computed from a part of the on-shell action \cite{Takayanagi:2011zk, Fujita:2011fp}. Using the conjectured correspondence between Islands and BCFTs \cite{Suzuki:2022xwv}, the boundary entropy can be obtained from a Ryu-Takayanagi surface whose tip is located at the boundary of the BCFT, $\phi=\pm\Delta\phi/2$. In our previous work, we reproduced this equivalence between the shadow entropy and the length of the minimal surface explicitly in the non-rotating BTZ black hole \cite{Kim:2023adq}. Meanwhile, we now have to consider an HRT surface for the rotating BTZ black hole. Thus, the thing is that which spacelike surface is selected for the minimal surface equivalent to this shadow entropy. To achieve the prescription, we will introduce a time coordinate that diagonalizes the metric and verify the equivalence in the next section.

As a comment on the shape of the EoW brane in a particular case, we examine the EoW brane configuration in the extremal black hole limit with $z_-\to z_+$. The EoW has a speed $\omega = z_+/z_-$, which is smaller than the unit speed of light. The maximum speed becomes 1 when the background is the extremal BTZ black hole. See (\ref{dG sol}) to get the limiting behavior. On the boundary of the black hole and near its horizon, one of the EoW branes behaves like
\begin{align}
&\Phi_+' \to \sqrt{1- \frac{z_+^2}{z_-^2}}\frac{\sigma}{\sqrt{1-\sigma^2}}\,,  &(\text{if}~~~z\to 0)\nonumber\\
&\Phi_+' \to \frac{\sigma}{1-\frac{z_+^2}{z_-^2}}\,. &(\text{if}~z\to z_+)
\end{align}
In addition, this EoW brane contacts the horizon at $\phi = \Delta\phi/2 \,+\,\Phi_+(z_+) $, where $\Phi_+(z_+)=z_+ \tanh^{-1}\sigma$. Thus, the extremal black hole gives rise to the following EoW brane behavior, like
\begin{align}
\Phi_+'(0) \to 0~~,~~\Phi_+'(z_+) \to \infty~~,~~\Phi_+ (z_+)= z_+ \tanh^{-1}\sigma \nonumber.
\end{align}
Therefore, the EoW brane is almost a straight line along the radial direction over most regions. Only near the horizon does the shape change, diving sharply into the point of contact.

\section{Thermodynamics of BCFT}

In this section, we derive the first law of thermodynamics for the BCFT system dual to the rotating BTZ black hole with two EoW branes that produce shadow entropies. Each boundary carries a tension parameter related to the boundary entropy. We also obtain HRT surfaces dual to the boundary entropies that are equivalent to the shadow entropies.

\subsection{Effective JT system for induced geometry on EoW brane}

We start with the induced metric on the EoW branes. The general form (\ref{ind_met shift lapse}) can be rewritten as the following intuitive expression:
\begin{align}\label{Induced Metric}
ds_{ind}^2 =h_{ab}d x^a d x^b = \frac{L^2}{z^2}
\left[-f(z)d\tilde{t}^2 +\frac{dz^2}{f(z)} +\left( \Phi'_\pm(z) dz  +   \frac{z_+}{z_-}\left(1-\frac{z^2}{z_+^2}\right)d\tilde{t} \right)^2\right]\,.
\end{align}
If the near-horizon limit ($z\to z_+$) from a local observer is taken, this metric approaches 
\begin{align}
\lim_{z\to z_+} ds_{ind}^2 =\lim_{z\to z_+} \frac{L^2}{z^2}\left[-f(z)d\tilde{t}^2 +\frac{dz^2}{f(z)} \right]\,.
\end{align}
Thus, this induced geometry has the same bulk horizon and temperature (\ref{T S}) as the BTZ black hole. Utilizing coordinate $z$, the temperature takes the following form:
\begin{align}\label{temp}
T = -\frac{f'(z_+)}{4\pi L}=\frac{1}{2\pi L\, z_+} \left(1-\frac{z_+^2}{  z_-^2}\right)\,.
\end{align}
In addition, the scalar curvature of the induced metric can be computed as
\begin{align}
R_{[h_{ab}]}= -\frac{2(1-\sigma^2)}{L^2}\,.
\end{align}
This constant negative curvature and the presence of a Killing horizon naturally identify the induced worldvolume theory as a JT black hole\footnote{In the present work, we relate the induced geometry to a JT system via a mapping. Still, we don't continue the discussion further. However, there is a more careful study on the emergence of semi-classical gravity in related braneworld setups. See \cite{Neuenfeld:2024gta} for the discussion.}. Here, one may naturally map the JT black hole system to the SYK model. Thus, this identification is particularly robust in the double-scaling limit of the SYK model, where the discrete chord diagrammatic structure approaches the continuous density of states characteristic of JT gravity \cite{Berkooz:2018jqr}. Crucially, the effective $AdS_2$ scale is renormalized by the brane tension, thereby allowing for a holographic interpretation of the boundary degrees of freedom. To construct the thermodynamic framework of the BCFT system, it would be desirable to study the first law of thermodynamics of the JT black hole, including the tension variation.

To derive the first law of thermodynamics, it is better to deal with a diagonal metric of the JT black hole based on a convenient time coordinate
$\hat{t}_i=\tilde{t}-\omega\,\Phi_i (z)/(1-\omega^2)$.
Here we have restored the brane index $i$ to emphasize that, for unequal tensions, $\Phi_i (z)$ depends on the individual tension $\sigma_i $. Consequently, although the two branes share the same rigid angular velocity $\omega=z_+/z_- $ within the solution branch considered above, the corresponding time coordinates \(\hat t_i\) are generally different when \(\sigma_1\neq\sigma_2\). 
For the explicit form of the induced metric below, we temporarily specialize to the equal-tension case,
$\sigma^1 = \sigma^2 =\sigma$, as considered in Sec.~2.2.
Using this coordinate and the junction equation (\ref{dG sol}), the induced metric on each brane takes the following form:
\begin{align}\label{metric 2d}
ds_{ind}^2=&-\frac{L^2}{z^2}\left(1-\frac{z^2}{z_-^2}\right) \left(1-\frac{z^2}{z_+^2}\right) \left(1-\omega^2\frac{\left(1-\frac{z^2}{z_+^2}\right)}{\left(1-\frac{z^2}{z_-^2}\right)}\right) d\hat{t}^2    \nonumber\\
&+\frac{L^2}{z^2}\frac{1}{\left(1-\frac{z^2}{z_-^2}\right) \left(1-\frac{z^2}{z_+^2}\right)} \left(1+\frac{\Phi'_{\pm}(z)^2 \left(1-\frac{z^2}{z_+^2}\right) \left(1-\frac{z^2}{z_-^2}\right)^2}{1-\omega^2}\right)dz^2 \nonumber\\
=&\frac{L^2}{z^2}\left[- \left(1-\frac{z^2}{z_+^2}\right) \left(1-\frac{z_+^2}{z_-^2}\right)d\hat{t}^2 + \frac{dz^2}{\left(1-\frac{z^2}{z_+^2}\right) \left(1-\sigma ^2+ \left(\frac{\sigma ^2}{z_+^2}-\frac{1}{z_-^2}\right)z^2\right)} \right].
\end{align}
In addition, let us take a further radial coordinate transformation given by
\begin{align}\label{r to w}
\frac{L^2}{z^2}=\frac{L^2 \left(\frac{1}{z_-^2}-\frac{\sigma ^2}{z_+^2}\right)}{1-\sigma ^2}+\frac{w^2}{1-\frac{z_+^2}{z_-^2}}\,.
\end{align}
This leads the induced metric to a familiar form of the JT black hole as
\begin{align}\label{metric jt ansatz}
ds_{ind}^2 = - A(w) L^2 d\hat{t}^2 + \frac{1}{A(w)} \frac{dw^2}{1-\sigma^2}\,,
\end{align}
where the metric function is
\begin{align}\label{sol A}
A(w) = \frac{w^2}{L^2}- b.
\end{align}
Here, the mass parameter $b$ of the JT black hole can be written in various ways as
\begin{align}
b=\frac{w_+^2}{L^2}=\frac{\left(1-\frac{z_+^2}{z_-^2}\right)^2}{\left(1-\sigma ^2\right) z_+^2} = \left( \frac{2\pi T L}{\sqrt{1-\sigma^2}} \right)^2.
\end{align}
From (\ref{metric jt ansatz}), the temperature can be read. The temperature is $T=A'(w_+)\sqrt{1-\sigma^2}/(4\pi)$. By construction, this temperature equals the bulk temperature (\ref{temp}).

Now, let us map the induced geometry (\ref{metric jt ansatz}) into a relevant 2-dimensional system, which can be proposed as follows:
\begin{align}\label{Action JT}
\mathcal{I}_{\text{JT}} = \frac{1}{2} \int d\hat{t}\,   dw \sqrt{-h}\,\mathsf{\Phi}\left( R_{\left[h_{ab}\right]} + \frac{2(1-\sigma^2)}{L^2}  \right)\,,
\end{align}
where $\mathsf{\Phi}$ is the dilaton field.
For unequal brane tensions, 
each EoW brane is described by a JT
system with its own effective AdS$_2$ radius
\begin{equation*}
L^i_{\rm eff} =\frac{L}{\sqrt{1-\left(\sigma^i \right)^2}}.
\end{equation*}
Thus, while the two EoW branes share the same rigid rotation in the bulk, their induced geometries correspond to JT black holes with different effective AdS$_2$ radii in general.
This action, employing the metric (\ref{metric jt ansatz}), leads to equations of motion given by
\begin{align}
L^2 A''(w) = 2\,,~L^2 A'(w) \mathsf{\Phi}'(w) = 2\mathsf{\Phi}(w)\,.
\end{align}
The dilaton $\mathsf{\Phi}$ can be selected by $\mathsf{\Phi}= w/L$ as a gauge choice. Then, this equation admits the metric function (\ref{sol A}) as a solution. Therefore, one may regard the induced metric as a geometry of the system whose cosmological constant is tension-dependent($\Lambda_{\text{JT}}=-(1-\sigma^2)/L^2$). In other words, this JT system has an effective curvature radius, $L_{\text{eff}}=L/\sqrt{1-\sigma^2}$.

Here, we leave a comment on the inner structure of the JT black hole. One may notice that an inner location $w_{-}$ corresponding to $z_{-}$ appears inside the horizon. From the explicit form of the EoW brane (\ref{dG sol}) or (\ref{EoW position}), one can notice that the EoW brane cannot exceed $z_-$ in the bulk. Therefore, the range of $w$ is from $w_{-}$ to $\infty$. However, the system (\ref{Action JT}) doesn't have such a limitation on the coordinate $w$. This is because the time coordinate transformation, between the bulk time $\tilde{t}$ and the time $\hat{t}$ of the JT system, becomes singular at $z=z_-$ or $w=w_-$. See the explicit form, 
\begin{align}
d\hat{t} = d\tilde{t} - \frac{\omega}{\left(1-\omega^2\right)}\Phi'_{\pm}(z)\,dz\,,
\end{align}
where $\Phi'_{\pm}(z)$ becomes divergent at $z=z_{-}$. Thus, the mapping of the induced metric to the JT system (\ref{Action JT}) is possible under the range, $w_-<w<\infty$, where $w_-$ is given by $w_-=\sigma \,w_+$. An EoW brane provides a geometric cutoff in the AdS bulk. When this serves as a Planck brane, it gives a UV cutoff in the context of AdS/CFT. Although there is no problem in the map from the induced geometry to the effective JT system (\ref{Action JT}) outside the horizon, it should be modified or consider a cutoff inside the horizon. In section \ref{Into the Unkonwn}, we deal with such an inner structure that plays the role of a junction (Joint) of a junction (EoW).

\subsection{The first law of thermodynamics for BCFT}

\begin{figure}
        \centering
        \begin{subfigure}{}
            \centering
            \includegraphics[width=120mm]{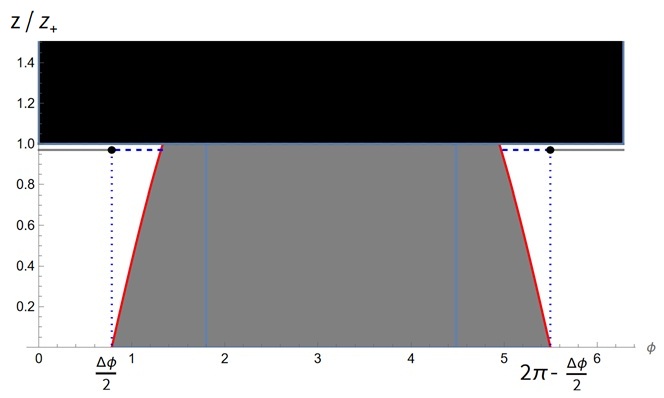}
        \end{subfigure}
                 \caption{{\bf A gravity dual to BCFT:} This is a cartoon of the BTZ black hole and two EoW branes at $t=0$. The black region denotes the interior part of the BTZ black hole. Two red curves show the EoW brane locations. The spatial volume of the BCFT is given by $\mathcal{V} = L \Delta\phi$. The solid gray lines near the horizon depict the BCFT bulk entropy. On the other hand, the dashed lines denote the shadow entropy generated by the EoW branes. Two EoW branes are moving along the $\phi$-direction with a constant speed $\omega = z_{+}/z_{-}$ under the time evolution.}\label{fig: Cartoon-shadow}
\end{figure}

Now, we are ready to discuss the thermodynamics of BCFT with two boundaries. A detailed configuration has been shown in Figure \ref{fig: Cartoon-shadow}. 
The two branes need not have identical tensions, and we denote the tension of the $i$-th brane by $\sigma^i$. As discussed above, unequal tensions give rise to different brane-adapted time coordinates $\hat{t}_i$ and hence to distinct effective JT geometries with $L_{\mathrm{eff}}^i=L/\sqrt{1-(\sigma^i)^2}$.
We therefore treat the thermodynamics of the two induced JT systems independently, while they remain coupled to the same rotating BTZ bulk.

Firstly, we focus on the thermodynamics of the $i$-th induced geometry. The mass, entropy, and temperature of the JT black hole are obtained as
\begin{align}
M_{\text{JT}}^i= \frac{\sqrt{1-(\sigma^{i})^2}}{2L} \left(\frac{w^i_+}{L}\right)^2~,~~S_{\text{JT}} =2\pi \mathsf{\Phi}(w_+) =2\pi \frac{w^i_+}{L}~,~~T =\frac{w^i_+}{2\pi L}\frac{\sqrt{1-(\sigma^{i})^2}}{L} .
\end{align}
Using these expressions, one can derive a variational relation among the quantities above; it then turns out that
\begin{align}\label{JT first law 00}
\delta M_{\text{JT}}^i - T \delta S_{\text{JT}}^i = -\frac{b^i}{2L}\frac{\sigma^i}{\sqrt{1-(\sigma^{i})^2}}\delta\sigma^i\,.
\end{align}
The left-hand side is the form of the usual first law of the JT black hole. If variation of the tension is not allowed, this relation is reduced to the usual thermodynamic law. This leads to an extended first law in which the variation of the tension is dual to that of a varying effective cosmological constant. This framework, based on (\ref{Action JT}), allows us to employ the `black hole chemistry' dictionary, elevating the tension to a thermodynamic pressure term with a well-defined conjugate volume \cite{Kubiznak:2012wp, Kubiznak:2016qmn, Arenas-Henriquez:2022www, Frassino:2022zaz, Arenas-Henriquez:2023hur}. See \cite{Mann:2025xrb} for a recent review on this topic. The application of these chemical variables to lower-dimensional gravity systems has demonstrated that the variation of the curvature radius is naturally conjugate to a thermodynamic volume \cite{Frassino:2015oca}.

In this approach, one should define a new kind of pressure that is negatively proportional to the cosmological constant. Thus, we will use the thermodynamic pressure associated with the JT black hole defined as follows:
\begin{align}
&\mathcal{P}^i_{\text{JT}} =-\frac{\Lambda^i_{\text{JT}}}{2}=\frac{1-(\sigma^{i})^2}{L^2}.
\end{align} 
Then, using (\ref{JT first law 00}), one can find the conjugate variable, which is called the `thermodynamic volume':
\begin{align}
\mathcal{V}_{\text{JT}}^i = \frac{b^i L}{4\sqrt{1-(\sigma^{i})^2}}\,.
\end{align}
The above quantities obey the following thermodynamic first law:
\begin{align}\label{first VdP}
\delta M_{\text{JT}}^i - T \delta S_{JT}^{i} = \mathcal{V}_{\text{JT}}^i \delta\mathcal{P}_{\text{JT}}^i\,.
\end{align}
This can be rewritten by $\delta U^i_{\text{JT}}= T \delta S_{JT}^{i} - \mathcal{P}_{\text{JT}}^i \delta\mathcal{V}_{\text{JT}}^i$, where the internal energy is given by $U_{\text{JT}}^i=M_{\text{JT}}^i - \mathcal{P}_{\text{JT}}^i \mathcal{V}_{\text{JT}}^i$. Therefore, the black hole mass is an enthalpy\footnote{Often the thermodynamic volume is expressed by an interior integration of the black hole as $\mathcal{V}_{\text{JT}}^i=\frac{1}{2 L^2}\int_0^{w^i_+} dw \sqrt{-h^i}\, \mathsf{\Phi}$. This formal expression suggests some relation to the complexity in quantum information theory. See \cite{Mann:2025xrb} for a recent summary of this research topic. In our case, the EoW branes cannot reach beyond the inner horizon $z_-$. This integral expression should be modified or used only formally.}.

Now, we suggest another interpretation of (\ref{JT first law 00}). In fact, the JT black hole is dual to the  SYK model \cite{Sachdev:1992fk, Kitaev:2015, Maldacena:2016hyu}. The duality relation between two models is elaborated in,{\it e.g.}, \cite{Maldacena:2016hyu, Maldacena:2016upp}. The current thermodynamic pressure $\mathcal{P}_{\text{JT}}$ is given by the effective $AdS_2$ radius $L^i_{\text{eff}}=L/\sqrt{1-(\sigma^i)^2}$. The inverse of curvature radius commonly corresponds to the averaged random coupling $\mathcal{J}_{\text{SYK}}$ of the SYK model. Therefore, it is legitimate to interpret the physical meaning of the thermodynamic pressure $\mathcal{P}_{\text{JT}}$ as an ensemble average of the squared coupling, $\mathcal{P}_{\text{JT}}\sim\mathcal{J}_{\text{SYK}}^2$.

For this reason, one may also construct the first law without using the thermodynamic volume. Using this parameter $\mathcal{J}_{\text{SYK}}$ dual to the inverse curvature radius, the variation (\ref{JT first law 00}) becomes
\begin{align}\label{first law JT SYK}
\delta M_{\text{JT}}^i - T \delta S_{\text{JT}}^i  =  \Theta^i_{\text{SYK}} \delta  \mathcal{J}_{\text{SYK}}^i\,,
\end{align} 
where the conjugate variable to the coupling turns out to be $\Theta_{\text{SYK}}^i \equiv {M_{\text{JT}}^i}/{\mathcal{J}^i_{\text{SYK}}}$. If we take a Legendre transformation, the above variation is equivalent to an interesting relation as $T \delta S_{\text{JT}}^i =\mathcal{J}_{\text{SYK}}^i \delta \Theta^i_{\text{SYK}}$. The mass and entropy correspond to the energy and logarithm of the density of states of the SYK model. The dual conjugate quantity  $\Theta_{\text{SYK}}^i$ is the black hole mass divided by the coupling. This is quite plausible, since the Hamiltonian of the SYK model is in such a form. Thus, we describe the thermodynamics in terms of the parameter $\mathcal{J}^i_{\text{SYK}}$, which corresponds to $1/L^i_{\text{eff}}$.

Let us find the first law of the total BCFT system illustrated in Figure \ref{fig: Cartoon-shadow}. The gray part is the truncated regions by the EoW branes, so the horizon area consists of the lengths of the solid and dashed lines near the horizon. The solid line denotes the entropy of a fan-shaped BTZ black hole. We have already displayed such an entropy $S_{\mathcal{V}}$ in section \ref{sec: BTZ thermodynamics}. On the other hand, the additional area from the dashed lines gives rise to the shadow entropy.

As we mentioned, the BCFT, in fact, has a profound relation with the Island formula\cite{Suzuki:2022xwv}. Through this Island/BCFT correspondence, a prescription for the boundary entropy was proposed. This boundary entropy is identified with the length of a minimal surface truncated by an EoW brane. The tip of this minimal surface lies floating on the boundary of the BCFT. It is also known that this segment of the minimal surface has constant length as the tip height ($z=z_*$) varies in the non-rotating BTZ black hole. Thus, we can move the tip height close to the horizon. Then, the part of the minimal surface coincides with the EoW shadow on the horizon. We provided a detailed explanation in the Appendix of our previous work \cite{Kim:2023adq}.

While the equivalence between shadow and boundary entropy is well established for static cases, the rotating BTZ background necessitates a covariant HRT prescription \cite{Hubeny:2007xt}. Our analysis demonstrates that this geometric identification remains robust even in the presence of angular momentum, preserving the Island/BCFT correspondence. We will return to this issue in the next subsection and propose HRT surfaces that give rise to the shadow entropy. So we regard the total horizon area in Figure \ref{fig: Cartoon-shadow} as the total entropy. Thus, we define the total entropy as $S_{total}\equiv S_\mathcal{V} + \sum_{i=1}^2 S_{shd}^i$, where $S_{shd}^i$ is given by (\ref{shadow entropy}). Using this total entropy, we can extend the first law (\ref{first law without boundary}) by grafting the first law of the JT black hole as follows:
\begin{align}\label{first-law01}
\delta\mathcal{E}_{\mathcal{V}} - T\delta S_{total} -\Omega \delta J_{\mathcal{V}} + \mathcal{P}\delta\mathcal{V} = \sum_{i=1}^2 \mathsf{w}^i \left(\delta M_{\text{JT}}^i - T \delta S_{JT}^i \right),
\end{align}
where the wrapping factor $\mathsf{w}^i$ is given by
\begin{align}\label{wrapping factor}
\mathsf{w}^i = \frac{L}{4\pi G} \frac{z_-^2 z_+}{z_-^2-z_+^2}\frac{\sqrt{1-(\sigma^i)^2}}{\sigma^i}\,.
\end{align}
We call the variation (\ref{first-law01}) the grafted first law of thermodynamics, which is the most general variation for the white region of the cartoon in Figure \ref{fig: Cartoon-shadow}. Notably, one more term than the previous result in \cite{Kim:2023adq} appears as the momentum flow or the angular momentum contribution, $\Omega \delta J_\mathcal{V}$.

This grafted-first law describes whatever parameter variations for this gravity dual to the BCFT. However, it is not a standard form of the first law. To make it a standard form, we rewrite the right-hand side in terms of the thermodynamic volume and pressure as $\sum_{i=1}^2 \mathsf{w}^i \mathcal{V}_{\text{JT}}^i \,\delta\mathcal{P}_{\text{JT}}^i$. Since the pressure has a well-defined physical meaning as the coupling of the SYK model, we promote it to the thermodynamic pressure defined in the BCFT system without any change. So, we define
\begin{align}
\mathcal{P}_{th}^i \equiv \mathcal{P}_{\text{JT}}^i\,.
\end{align}
Now, the conjugate thermodynamic variable to this pressure, by considering the grafted thermodynamics, is as follows:
\begin{align}
\mathcal{V}_{th}^i \equiv \mathsf{w}^i \mathcal{V}_{\text{JT}}^i\,.
\end{align}
Then the grafted-first law can be rewritten as
\begin{align}
\delta\mathcal{E}_{\mathcal{V}} - T\delta S_{total} -\Omega \delta J_{\mathcal{V}} + \mathcal{P}\delta\mathcal{V} = \sum_{i=1}^2 \mathcal{V}_{th}^i \delta\mathcal{P}_{th}^i\,.
\end{align}
In addition, we define the internal energy as follows:
\begin{align}\label{Internal energy}
U \equiv \mathcal{E}_{\mathcal{V}} - \sum_{i=1}^2 \mathcal{P}_{th}^i \mathcal{V}_{th}^i\,.
\end{align}
Then we can finally obtain the standard form of the first law as 
\begin{align}\label{U PV}
\delta U = T\delta S_{total} + \Omega \delta J_{\mathcal{V}} - \mathcal{P}\delta\mathcal{V}- \sum_{i=1}^2\mathcal{P}_{th}^i \delta\mathcal{V}^i_{th}\,.
\end{align}
From (\ref{Internal energy}), one can see that the original black hole mass $ \mathcal{E}_{\mathcal{V}} = U + \sum_{i=1}^2 \mathcal{P}_{th}^i \mathcal{V}_{th}^i $ turns out to be the enthalpy. Looking at the internal energy (\ref{Internal energy}), the last term is the contribution from the two boundaries.

As a final remark, we provide another form of the first law in terms of the averaged coupling of the dual SYK model. The mass parameter $\mathcal{E}_{\mathcal{V}}$ of the BTZ black hole or the BCFT bulk energy varies in the following form:
\begin{align}
\delta\mathcal{E}_{\mathcal{V}} - T\delta S_{total} -\Omega \delta J_{\mathcal{V}} + \mathcal{P}\delta\mathcal{V} = \sum_{i=1}^2 \mathsf{w}^i \Theta_{\text{SYK}}^i \delta \mathcal{J}_{\text{SYK}}^i\,.
\end{align}
The conjugate variable to the coupling $\mathcal{J}_{\text{SYK}}$ is now defined by $\Theta^i\equiv \mathsf{w}^i \Theta_{\text{SYK}}^i$. This can be written in terms of the coupling and temperature\footnote{ We provide explicitly related expressions as $\Theta_{\text{SYK}}^i=\frac{2\pi^2 T^2}{\mathcal{J}^i_{\text{SYK}}{}^2}$ and $\mathsf{w}^i=\frac{L}{8\pi^2 G}\frac{\mathcal{J}^i_{\text{SYK}}}{\sqrt{1- \mathcal{J}^i_{\text{SYK} }{}^2 L^2 }}\frac{1}{T}$.}:
\begin{align}
\Theta^i = \frac{L}{4G}\frac{T}{\mathcal{J}^i_{\text{SYK}}\sqrt{1-\mathcal{J}^i_{\text{SYK}}{}^2 L^2  }}\,.
\end{align}
Then another form of the first law can be written as follows:
\begin{align}\label{first law final}
\delta U = T\delta S_{total} + \Omega \delta J_{\mathcal{V}} - \mathcal{P}\delta\mathcal{V} -\sum_{i=1}^2 \mathcal{J}^i_{\text{SYK}}\delta\Theta^i\,
\end{align}
where $U=\mathcal{E}_{\mathcal{V}}- \sum_{i=1}^2\Theta^i \mathcal{J}^i_{\text{SYK}}$, which is equivalent to (\ref{Internal energy}). The boundary SYK effect negatively contributes to this internal energy, like the previous expression (\ref{Internal energy}). Therefore, (\ref{first law final}) is the expression of the first law in terms of the bulk CFT and boundary effective SYK degrees of freedom.

Here, note that the boundary effect lowers the internal energy $U$. This reflects an attractive interaction between the boundary and bulk degrees of freedom. In addition, this interaction energy is proportional to $- \sum_{i} (T/\sigma^i)$. 
At first sight, this behavior appears singular in the limit $\sigma^i\to0 $. However, this limit makes the EoW branes disappear, and the other bulk, erased by the EoW branes, revives and is attached to the BCFT volume. This is an artifact of the junction-probe formulation. Therefore, we conjecture that the stability of this system is maintained in a reasonable parameter region for a nonvanishing tension, even though we can not put forward a strict strategy. However, a more concrete analysis is needed for the thermodynamic BCFT  system. We leave this as a future study.

In summary, we found the first law of thermodynamics for the induced geometry as (\ref{first VdP}) and (\ref{first law JT SYK}). The former was already introduced in our previous work. In addition, (\ref{U PV}) and (\ref{first law final}) are the general first law for the BCFT system, including the momentum flow and boundary degrees of freedom.

\subsection{Minimal surface and shadow entropy}

Now, we would like to find HRT surfaces that give rise to the boundary entropy through the Island/BCFT correspondence. We will also show that the boundary entropy obtained in this way equals the shadow entropy.

The covariant formulation of minimal surfaces was studied in \cite{Hubeny:2007xt}. A minimal surface in a non-static geometry can be found as an intersection of the past and future light cones for the boundary wedges. Even though this approach is robust for finding minimal surfaces, there is an easier way to do so in the rotating BTZ background. In the rotating BTZ black hole, one may use a coordinate transformation to simplify the metric. To do this, it is useful to employ a diagonalized metric by using two coordinates given by
\begin{align}
\varphi = \phi - \omega\,\tilde{t}~~,~~\bar{t} = \tilde{t} -\frac{\omega}{1-\omega^2} \, \varphi  = \frac{\tilde{t}-\omega\,\phi}{1-\omega^2}\,.
\end{align}
Then, the metric takes the following form in terms of these coordinates: 
\begin{align}\label{metric_diagon 00}
ds^2 = \frac{L^2}{z^2}\left[  -\left(1-\frac{z^2}{z_+^2}\right) \left(1-\frac{z_+^2}{z_-^2}  \right)d\bar{t}^2 + \frac{ \left(1-\frac{z^2}{z_-^2}\right)}{\left(1-\frac{z_+^2}{z_-^2}\right)} d{\varphi}^2 +\frac{{dz}^2}{\left(1-\frac{z^2}{z_-^2}\right) \left(1-\frac{z^2}{z_+^2}\right)} \right]\,.
\end{align}
Taking further transformation as
\begin{align}
&\mathcal{X} =\sqrt{\frac{1-\frac{z^2}{z_+^2}}{1-\frac{z^2}{z_-^2}}} e^{-\frac{\varphi }{z_+}}\cosh \left[\left(1-\frac{z_+^2}{z_-^2}\right)\frac{ \bar{t}}{z_+}\right]\,.\nonumber\\
&\mathcal{T} =\sqrt{\frac{1-\frac{z^2}{z_+^2}}{1-\frac{z^2}{z_-^2}}} e^{-\frac{\varphi }{z_+}}\sinh \left[\left(1-\frac{z_+^2}{z_-^2}\right)\frac{ \bar{t}}{z_+}\right]\,,\nonumber\\
&\mathcal{Z}=\sqrt{\frac{{\frac{z_-^2}{z_+^2}-1}}{{\frac{z_-^2}{z^2}-1}}} e^{-\frac{\varphi }{z_+}}\,,
\end{align}
the metric (\ref{metric_diagon 00}) changes to the $\text{Poincar\'e}$ metric,
\begin{align}
ds^2 = \frac{L^2}{\mathcal{Z}^2}\left(-d\mathcal{T}^2 + d\mathcal{X}^2 + d\mathcal{Z}^2 \right)\,.
\end{align}
This is the well-known metric, and one can easily find minimal surfaces or EoW branes.

\begin{figure}
        \centering
        \begin{subfigure}{}
            \centering
            \includegraphics[width=80mm]{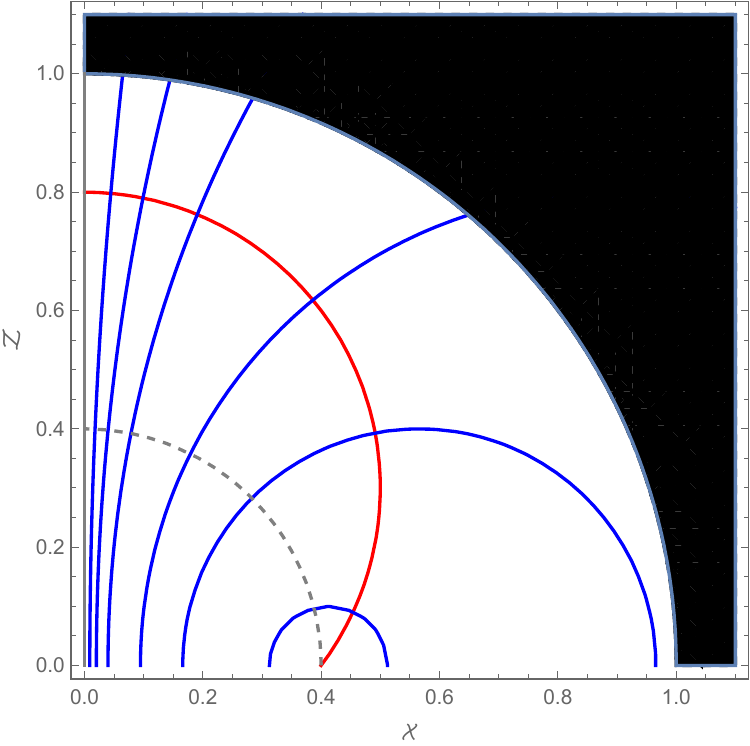}
        \end{subfigure}
                 \caption{{\bf Minimal surfaces and a EoW brane at $\mathcal{T}=0$:} The red curve denotes an EoW brane, and the gray-dashed curve depicts a constant $\varphi$ line. The blue curves are minimal surfaces, whose tips lie on the dashed curve. Thus, the boundary entropy is given by the lengths of the blue curves between the EoW brane and the dashed line. These lengths are indeed all equal, so the shadow entropy, given by the length from the dashed line to the red curve at the horizon $\mathcal{X}=0$, is also equal.}\label{fig: Cartoon-quarter circle}
\end{figure}

Let's focus on $\mathcal{T}=0$. This case is achieved by the vanishing time coordinate $\bar{t}$. Then, the other two coordinates become
\begin{align}
\mathcal{X} =\sqrt{\frac{1-\frac{z^2}{z_+^2}}{1-\frac{z^2}{z_-^2}}} e^{-\frac{\varphi }{z_+}}~,~\mathcal{Z}=\sqrt{\frac{{\frac{z_-^2}{z_+^2}-1}}{{\frac{z_-^2}{z^2}-1}}} e^{-\frac{\varphi }{z_+}}\,.
\end{align} 
Then, we can invert the previous coordinates as follows:
\begin{align}
\varphi = \frac{z_+}{2}\log \left(\frac{1}{\mathcal{X}^2+\mathcal{Z}^2} \right)~,~\frac{z}{z_+}=\frac{\mathcal{Z}}{\sqrt{\mathcal{X}^2 \left(1-\frac{z_+^2}{z_-^2}\right)+\mathcal{Z}^2}}\,.
\end{align}
The boundary of the AdS space and the horizon are located at $\mathcal{Z}=0$ and $\mathcal{X}=0$. If we restrict the coordinate $\varphi$ to positive values, the full space can be confined to the quarter unit circle in the positive $\mathcal{X}$ and $\mathcal{Z}$ region, as shown in Figure \ref{fig: Cartoon-quarter circle}.

A minimal surface is given by $(\mathcal{X}-\mathcal{X}_0)^2 + \mathcal{Z}^2= r_m^2$, where ($\mathcal{X}_0^2>r_m^2$)\footnote{This minimal surface solution is a representative solution up to a boosting $\mathcal{T}$ and $\mathcal{X}$. One may consider further transformation with $\mathcal{X}+\mathcal{T} \to e^{\xi}(\mathcal{X}+\mathcal{T})$ and $\mathcal{X}-\mathcal{T} \to e^{-\xi}(\mathcal{X}-\mathcal{T})$. However, we restrict our consideration to only $\mathcal{T}=0$ in the present work. }. In addition, an EoW brane is described by $\mathcal{X}^2 + (\mathcal{Z}-\mathcal{Z}_0)^2= r_\sigma^2$ with a condition ($r_\sigma^2 >\mathcal{Z}_0^2$). We show an example in Figure \ref{fig: Cartoon-quarter circle}. In the figure, the shadow entropy is obtained by taking the length of a straight line at $\mathcal{X}=0$ between the red and gray-dashed circles. The red circle denotes an EoW brane, and the gray-dashed circle depicts the constant $\varphi$.

On the other hand, the blue curves denote minimal surfaces whose tips lie on the gray dashed curve. Therefore, the boundary entropy is given by the lengths of the blue curves between the dashed and the red circles. If all blue curves give rise to the same boundary entropy, then we may move the tip to the horizon. This proves the equivalence of the shadow entropy and the boundary entropy. This equivalence for the blue lines can also be shown in the $(\bar{t}, \varphi, z)$-coordinate system through an explicit calculation.

We start with the diagonal metric (\ref{metric_diagon 00}) to show the equivalence between the shadow and boundary entropies. We will consider a codimension 2 surface with a fixed $\bar{t}$. For instances, $\bar{t}=0$ surface means $\tilde{t}-\omega\,\phi = 0$, which is a spacelike surface. Taking into account a cylindrical coordinate system $(e^{-z/z_+}, \phi, \tilde{t})$, this spacelike surface is a spiral sheet in the coordinate cylinder. We will keep our discussion on these spiral sheets. To find a relevant codimension 2 surface rotating rigidly, we restrict $d\bar{t}=0$ and $z=Z_{ms}(\varphi)$ to the metric (\ref{metric_diagon 00}), and construct the area functional given by
\begin{align}
\mathcal{S}_b &= \frac{\mathcal{A}}{4G} = \frac{L}{4G}\int d\varphi \frac{1}{Z_{ms}(\varphi )}\sqrt{\frac{Z_{ms}'(\varphi )^2}{\left(1-\frac{Z_{ms}(\varphi )^2}{z_-^2}\right) \left(1-\frac{Z_{ms}(\varphi )^2}{z_+^2}\right)}+\frac{1-\frac{Z_{ms}(\varphi )^2}{z_-^2}}{1-\frac{z_+^2}{z_-^2}}}\,.
\end{align}
Using a conserved quantity about $\varphi$-independence of this effective Lagrangian, one can find the equation of motion for the minimal surface as
\begin{align}
Z_{ms}'(\varphi)= - \frac{\left(1-\frac{Z_{ms}(\varphi )^2}{z_-^2}\right) \sqrt{1-\frac{Z_{ms}(\varphi )^2}{z_+^2}} \sqrt{\frac{z_*^2}{Z_{ms}(\varphi )^2}-1}}{\sqrt{1-\omega^2} \sqrt{1-\frac{z_*^2}{z_-^2}}}\,.
\end{align}
This minimal surface is anchored at the boundary of the bulk geometry. In the above, $z_*$ denotes the maximum value of $Z_{ms}(\varphi_*)$, where $Z_{ms}'(\varphi_*)=0$ to have a smooth configuration. Adopting this boundary condition, one can find the minimal surface solution as follows:
\begin{align}
z=Z_{ms}(\varphi)= z_* \sqrt{\frac{\left(1-\frac{z_+^2}{z_-^2}\right)-\frac{z_+^2}{z_*^2} \left(1-\frac{z_*^2}{z_-^2}\right) \tanh ^2\left(\frac{\phi -\frac{\Delta \varphi }{2}}{z_+}\right)}{\left(1-\frac{z_+^2}{z_-^2}\right)-\left(1-\frac{z_*^2}{z_-^2}\right) \tanh ^2\left(\frac{\phi -\frac{\Delta \varphi }{2}}{z_+}\right)}}\,.
\end{align}
This solution and the EoW brane curve $\Phi_+(z)$ intersect at $z=z_Q$, where
\begin{align}
z_Q=z_*\frac{\sqrt{1-\sigma ^2} }{\sqrt{1-\frac{ z_*^2}{z_+^2}\sigma^2}}\,.
\end{align}
This intersection and the equation of motion give the boundary entropy $\mathcal{S}_b$ as follows:
\begin{align}
\mathcal{S}_b =&\frac{L}{4G} \int_{z_Q}^{z_*} dZ_{ms} \frac{z_+z_*}{Z_{ms}\sqrt{z_+^2-Z_{ms}^2}\sqrt{z_*^2-Z_{ms}^2}}\nonumber\\
=&\frac{L}{4G} \int_{\sqrt{\frac{1-\sigma^2}{1-\eta^2\sigma^2}}}^1 \,dx\left[ \frac{1}{x\sqrt{1-x^2}\sqrt{1-\eta^2x^2}}  \right] \nonumber\\
=&\frac{L}{4G}\int_0^\sigma \frac{du}{1-u^2}= \frac{L}{4G} \tanh^{-1}\sigma \,,
\end{align}
where we used a change of variable $u =\sqrt{(1-x^2)/(1-\alpha_*^2 x^2)}$ with $\alpha_*=z_*/z_+$. This result implies that the tip position $z_*$ does not affect the boundary entropy, so one may take $z_*\to z_+$. Therefore, the boundary entropy $\mathcal{S}_b$ is same with the shadow entropy $S_{shd}$ and the total entropy is given by
\begin{align}
S_{total} = S_{\mathcal{V}} +\sum_{i=1}^2 \mathcal{S}_b^i\,,
\end{align}
where $\mathcal{S}_b^i$ is defined by a HRT surface on  a constant $\bar{t}$ surface for i-th EoW brane.

\section{Into the Interior World} \label{Into the Unkonwn}

\begin{figure}
        \centering
       \begin{subfigure}{}
            \centering
            \includegraphics[width=50mm]{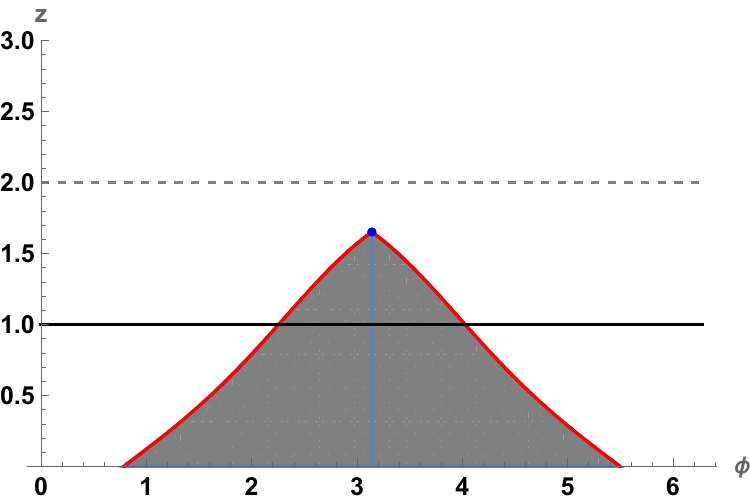}
         \end{subfigure}
          \centering
        \begin{subfigure}{}
            \centering
            \includegraphics[width=50mm]{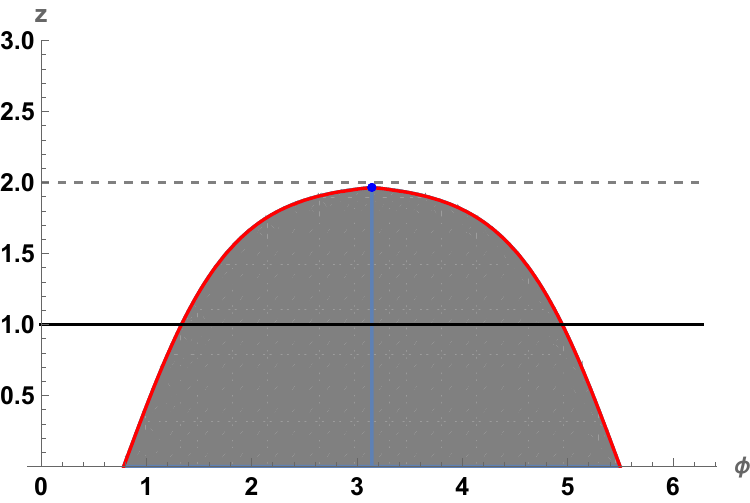}
        \end{subfigure}
         \begin{subfigure}{}
            \centering
            \includegraphics[width=50mm]{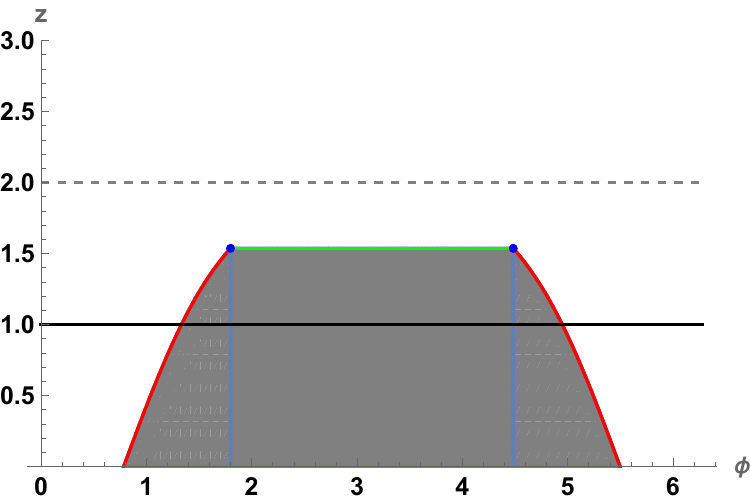}
         \end{subfigure} \\
        \begin{subfigure}{}
            \centering
            \includegraphics[width=50mm]{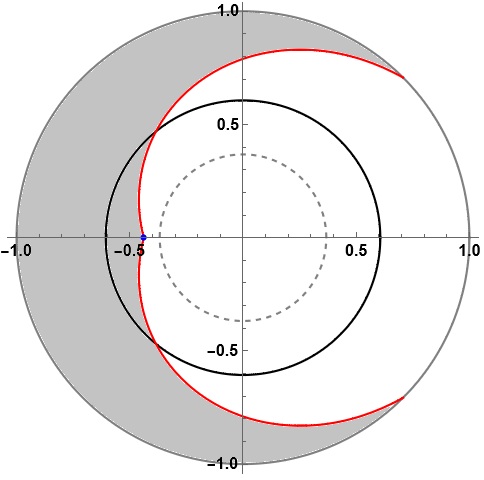}
         \end{subfigure}
         \begin{subfigure}{}
            \centering
            \includegraphics[width=50mm]{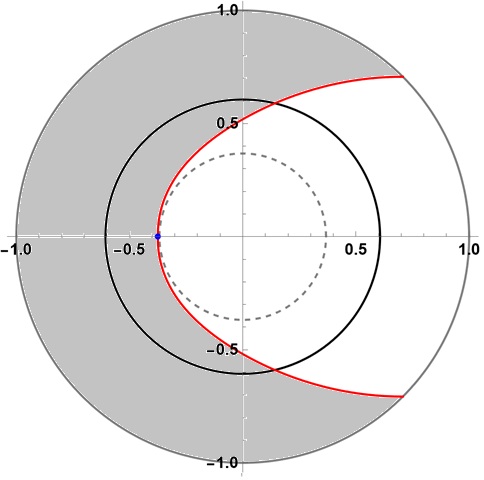}
        \end{subfigure}
         \begin{subfigure}{}
            \centering
            \includegraphics[width=50mm]{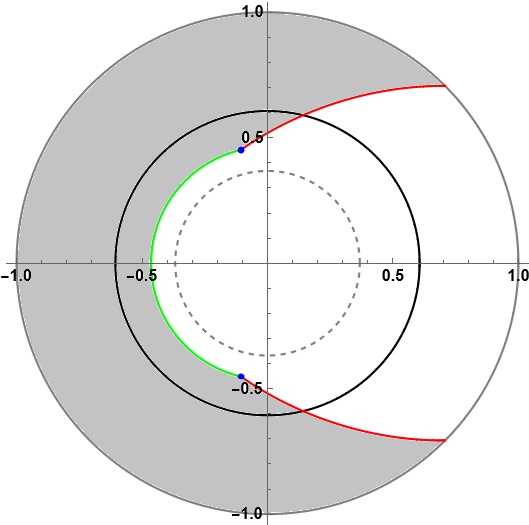}
         \end{subfigure}
         \caption{{\bf Two identical EoW branes in the BTZ black hole at $\tilde{t}=0$:} We set $z_+=1$, $z_-=2$, and $\sigma=0.5(\text{Left}),\, 0.9(\text{Middle, Right})$. The red curves and blue dots denote the EoW branes and the joints. The lower figures show the upper configurations in the polar coordinates defined by $(\phi,\, e^{-z/2})$. Two EoW branes in the first and second column figures meet at the blue dots outside the inner horizon (Dashed curve). The third column figures show a spacelike brane (Green lines) inside the horizon and two joints.}\label{fig: EoW cartoon01}
\end{figure}{}

\begin{figure}
        \centering
        \begin{subfigure}{}
            \centering
            \includegraphics[width=110mm]{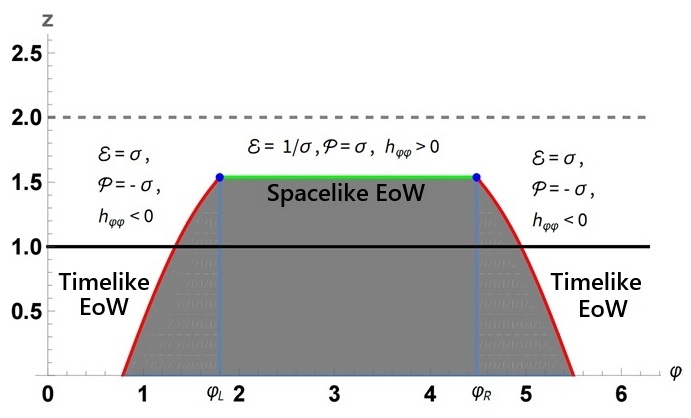}
        \end{subfigure}
                 \caption{{\bf An exmaple of double-joint EoW brane}  .}\label{fig: EoW cartoon02}
\end{figure}{}

In the previous section, we focused on the spacetime outside the horizon bordered by two EoW branes. These branes extend from the boundary of the AdS space into the black hole's interior. The solutions in (\ref{EoW position}) cross the horizon smoothly, notwithstanding signature changing in some metric components at the horizon. If the BCFT system has a well-defined and stable thermodynamics, the total bulk system enclosed by the EoW branes should show a definite configuration even inside the horizon. We try to clarify this issue and suggest candidates for the interior configuration in this section. In our previous work \cite{Kim:2023adq}, we addressed this issue for the non-rotating black hole using only the intersection scale of the interior brane configuration. Now, we will develop such a discussion on a more physical ground.

For the non-rotating BTZ black hole, two EoW branes extending outside to inside the horizon intersect at a single point in the interior. This intersection point behaves like a particle at rest, avoiding the curvature singularity. This point moves out by lowering the temperature, as studied in \cite{Kim:2023adq}. For the rotating BTZ configuration, the brane solution (\ref{EoW position}) tells us that an EoW brane cannot cross the inner horizon $z_-$ since $\Phi_{\pm}(z)$ becomes singular near $z=z_-$. To visualize the EoW brane configuration, we plot several shapes of the branes in Figure \ref{fig: EoW cartoon01}. The gray regions are domains truncated by the EoW branes. The black solid and gray dashed lines of the figures denote the outer and inner horizons, respectively. The figures depict initial configurations. The configurations in the lower figures rotate rigidly about the origin ($z=\infty$) with the angular velocity $\omega$ under time evolution, and the configurations in the upper figures move right with the same constant speed $\omega$.

\subsection{Single-joint EoW brane}

In this section, we study the first possible configuration within the black hole's horizon. As we did in the previous section, we use the diagonalized metric (\ref{metric_diagon 00}). In the interior, the metric can be written as
\begin{align}\label{metric_diagon}
ds^2=& -( e^z)^2 + (e^{\bar{t}})^2 + (e^{\varphi})^2\nonumber\\
=&\frac{L^2}{z^2}\left[ -\frac{{dz}^2}{\left(1-\frac{z^2}{z_-^2}\right) \left(\frac{z^2}{z_+^2}-1\right)}+   \left(\frac{z^2}{z_+^2}-1\right) \left(1-\frac{z_+^2}{z_-^2}  \right)d\bar{t}^2 + \frac{ \left(1-\frac{z^2}{z_-^2}\right)}{\left(1-\frac{z_+^2}{z_-^2}\right)} d{\varphi}^2 \right]\,,
\end{align}
where $e^z$, $e^{\bar{t}}$, and $e^\varphi$ are the dreibein one-forms.

Let us take the location of an EoW brane as $z=Z(\varphi)$ in terms of $\varphi=\phi-\omega \tilde{t}$ representing the rigid rotation. Furthermore, it is convenient to take a coordinate $\eta$ for an ADM decomposition as
\begin{align}
\epsilon_d\, \eta = z - Z(\varphi)\,,
\end{align}
where $\epsilon_d$ can be taken as $\pm 1$ for the choice of the orientation for the normal vector. We will determine this sign later. A normal vector $n_M dx^M$ to the EoW brane is proportional to $dz - Z'(\varphi)d\varphi$.  Note that $z_-> Z(\varphi) \geq z_+$ in the interior of the black hole. The bulk metric is decomposed by
\begin{align}
ds^2 =  ( \epsilon_s N^2+V_a V^a ) d\eta^2  + 2 V_a dx^a d\eta +  h_{ab}dx^a dx^b\,,
\end{align}
where $\epsilon_s = n_M n^M$, {\it i.e.}, $\epsilon_s = \mp 1$ for timelike or spacelike normal vector $n_M$ corresponding to spacelike or timelike EoW branes.

Under this ADM decomposition, the lapse function and the shift vector can be found as
\begin{align}
&N^2  = \frac{-\,\epsilon _s\left(1-\frac{z^2}{z_-^2}\right) }{z^2 \left[\left(1-\frac{z^2}{z_-^2}\right)^2 \left(\frac{z^2}{z_+^2}-1\right)-\left(1-\frac{z_+^2}{z_-^2}\right) Z'(\varphi )^2\right]}\,,\nonumber\\
&V_a dx^a  =  -\epsilon_d L^2 \frac{z_+^2}{z^4} \frac{Z'(\varphi )}{\left(1-\frac{z^2}{z_-^2}\right) \left(1-\frac{z_+^2}{z^2}\right)}  d\varphi\,.
\end{align}
Also, the induced metric is
\begin{align}
ds_{ind}^2 = \frac{L^2}{z^2} \left[ \left(\frac{z^2}{z_+^2}-1\right) \left(1-\frac{z_+^2}{z_-^2}\right) d\bar{t}^2  + \left(\frac{1-\frac{z^2}{z_-^2}}{1-\frac{z_+^2}{z_-^2}}-\frac{Z'(\varphi )^2}{\left(1-\frac{z^2}{z_-^2}\right) \left(\frac{z^2}{z_+^2}-1\right)}\right) d\varphi^2  \right]\,.
\end{align}
By using the above expressions, one can calculate the extrinsic curvature:
\begin{align}
K_{ab} =  \frac{1}{2N}\left( \partial_\eta h_{ab} - D^{[h]}_a V_b - D^{[h]}_b V_a  \right)\,,
\end{align}
where $h_{ab}$ is the induced metric. Then, the junction equation for an EoW brane is 
\begin{align}\label{Junct 00}
K h_{ab} -K_{ab}=  T^{(m)}_{ab} =  \frac{1}{L} \text{diag}\left[-\, \mathcal{E}(\varphi)h_{\bar{t}\bar{t}} , \mathcal{P}(\varphi)h_{\varphi\varphi} \, \right]\,.
\end{align}

By plugging the above ingredients into this junction equation, one can easily see that the off-diagonal components vanish. The $\varphi\varphi$-component of the junction equation can be written as 
\begin{align}\label{Junct 01}
\frac{-\epsilon_s\epsilon_d\mathcal{P}(\varphi )}{ \left(1-\frac{Z(\varphi )^2}{z_-^2}\right)} \sqrt{\epsilon_s\left[\frac{\left(1-\frac{z_+^2}{z_-^2}\right)}{\left(1-\frac{Z(\varphi )^2}{z_-^2}\right)} Z'(\varphi )^2-\left(1-\frac{Z(\varphi )^2}{z_-^2}\right) \left(\frac{Z(\varphi )^2}{z_+^2}-1\right)\right]}=1\,.
\end{align} 
As one can see, the interior of the square root and $\epsilon_s\epsilon_d(-\mathcal{P}(\varphi))$ should be positive to have a physical solution\footnote{This can also be written using the bulk metric as $(-\epsilon_s\epsilon_d\mathcal{P})  \frac{L^3g^{\varphi\varphi}}{z^3}\sqrt{\epsilon_s\left[g^{\varphi\varphi}Z'^2+g^{zz}\right]}|_{z=Z(\varphi)}=1-\frac{z_+^2}{z_-^2}$.}. Under this requirement and the expression of $Z'(\varphi)^2$, the $\bar{t}\bar{t}$-component of the junction equation takes the following form:
\begin{align}\label{Junct 02}
Z''(\varphi) =&- \frac{\epsilon _d z_+^2 }{Z(\varphi)^3} \frac{ \left(1-\frac{Z(\varphi )^2}{z_-^2}\right)^3}{\left(1-\frac{z_+^2}{z_-^2}\right) \left(1-\frac{z_+^2}{Z(\varphi )^2}\right)}\frac{\mathcal{E}(\varphi )}{|\mathcal{P}(\varphi )|^3}\nonumber\\
&-\frac{z_+^2 \epsilon _s}{\mathcal{P}(\varphi)^2 Z(\varphi )^3}\frac{\left(1-\frac{Z(\varphi )^2}{z_-^2}\right)^2 \left(1-\frac{3 Z(\varphi )^4}{z_-^2 z_+^2}+\frac{2 Z(\varphi )^2}{z_-^2}\right)}{\left(1-\frac{z_+^2}{z_-^2}\right) \left(\frac{z_+^2}{Z(\varphi )^2}-1\right)}\nonumber\\
&+\frac{Z(\varphi ) \left(1-\frac{Z(\varphi )^2}{z_-^2}\right)}{z_+^2 \left(1-\frac{z_+^2}{z_-^2}\right)} \left( 1 -\frac{3 Z(\varphi )^2}{z_-^2}+\frac{2 z_+^2}{z_-^2}\right)\,.
\end{align} 
Then, we are ready to obtain solutions for interior EoW brane configurations by solving (\ref{Junct 01}) and (\ref{Junct 02})\footnote{Using the junction equations (\ref{Junct 01},\ref{Junct 02}), one may obtain the conservation law given by
\begin{align}\label{Conservation 00}
\mathcal{P}'(\varphi ) =- \frac{\mathcal{E}(\varphi )+\mathcal{P}(\varphi )}{\frac{Z(\varphi )^2}{z_+^2}-1} \frac{Z'(\varphi )}{Z(\varphi )}\,.
\end{align}}.

Now, let us reproduce the EoW branes obtained in the previous section. To describe this spacelike normal vector case, we choose $\epsilon_s =1$. In addition, we set the energy and pressure on the brane as $\mathcal{E}= -\mathcal{P}=\sigma$, the same as outside of the horizon. Then, we have to choose $\epsilon_d=1$. By plugging $\mathcal{P}=-\sigma$, the $\varphi\varphi$-component (\ref{Junct 01}) takes the following form:
\begin{align}\label{dZ eq 00}
Z'(\varphi) = \pm \frac{\left(1-\frac{Z(\varphi )^2}{z_-^2}\right) \sqrt{1 - \frac{Z(\varphi)^2}{z_-^2}+ \sigma^2   \left(\frac{Z(\varphi )^2 }{z_+^2}-1\right)  }}{\sigma  \sqrt{1-\frac{z_+^2}{z_-^2}}}\,.
\end{align}
This reproduces (\ref{dG sol}) since $Z'(\varphi)\Phi'(z)=1$. In addition, this equation admits the following solutions:
\begin{align}
Z(\varphi)= \left\{
\begin{array}{c}
  \frac{z_+\sqrt{1-\sigma ^2} \tanh \left(\frac{\varphi -\frac{\Delta \varphi }{2}}{z_+}\right)}{\sqrt{\sigma ^2 \left(1-\frac{z_+^2}{z_-^2}\right)+\left(\frac{z_+^2}{z_-^2}-\sigma ^2\right) \tanh ^2\left(\frac{\varphi -\frac{\Delta \varphi }{2}}{z_+}\right)}}\,~~~~\left(\frac{\Delta\varphi}{2}\leq\varphi\leq\pi\right) \\
 \frac{z_+\sqrt{1-\sigma ^2} \tanh \left(\frac{2\pi-\varphi -\frac{\Delta \varphi }{2}}{z_+}\right)}{\sqrt{\sigma ^2 \left(1-\frac{z_+^2}{z_-^2}\right)+\left(\frac{z_+^2}{z_-^2}-\sigma ^2\right) \tanh ^2\left(\frac{2\pi-\varphi -\frac{\Delta \varphi }{2}}{z_+}\right)}}\,~~~~~~\left(\pi<\varphi\leq\pi+\frac{\Delta\varphi}{2}\right) \\
\end{array}
\right.\,.
\end{align}
This is nothing but the inverse function of (\ref{EoW position}). The solution outside the horizon can obviously extend to the interior region.

Let us look at the first and second columns in Figure \ref{fig: EoW cartoon01}. As the figures show, one EoW brane contacts with the other one at $\varphi=\pi$. This configuration can be regarded as a single-brane configuration with a joint or cusp. To physically realize such a joint, one may consider highly distributed energy at the joint position. This highly localized energy can bend the EoW brane, forming a cusp.  Also, the junction equation (\ref{Junct 00}) can admit this cusp as a physical configuration. We would now like to construct this nonsmooth junction configuration as a physical object. This joint configuration, or the junction of the junction, has been studied in a gravity system \cite{Hayward:1993my}. We use this theoretical ingredient to introduce the joint.

To find the energy distribution associated with the joint, we will consider (\ref{Junct 01}) and the $\bar{t}\bar{t}$-component of the junction equation (\ref{Junct 00}) rather than the conservation law (\ref{Conservation 00}). (\ref{Junct 01}) can be solved by (\ref{dZ eq 00}) with replacing $\sigma$ with $-\mathcal{P}(\varphi)$. Using this equation, we take an integration near the joint at $\varphi=\pi$, given by
\begin{align}
\int_{\pi - \tilde{\delta} }^{\pi+\tilde{\delta}}d\varphi Z'(\varphi)= Z(\pi + \tilde{\delta}) -Z(\pi - \tilde{\delta}).
\end{align}
Here, $\tilde{\delta}$ is a tiny angle. We require a continuous EoW brane, {\it i.e.}, $\lim_{\tilde{\delta}\to 0}[Z(\pi + \tilde{\delta}) -Z(\pi - \tilde{\delta})]=0$. Therefore, the $\mathcal{P}$ is the constant, given by the tension, over this tiny range without a singular pressure distribution.

Now, let us consider the other junction equation (\ref{Junct 02}). The equation has the following structure:
\begin{align}
Z''(\varphi)=-\frac{z_+^2 \mathcal{E}(\varphi ) \left(1-\frac{Z(\varphi )^2}{z_-^2}\right)^3}{\sigma ^3 \left(1-\frac{z_+^2}{z_-^2}\right) Z(\varphi )^3 \left(1-\frac{z_+^2}{Z(\varphi )^2}\right)}+\tilde{\mathbb{F}}(Z(\varphi),\sigma,z_+,z_-)\,,
\end{align} 
where we used (\ref{dZ eq 00}). Using this equation, we may again consider the integration near $\varphi=\pi$ as
\begin{align}
\int_{\pi-\tilde{\delta}}^{\pi + \tilde{\delta}}d\varphi Z''(\varphi)= Z(\pi + \tilde{\delta}) -Z(\pi - \tilde{\delta})=-2 Z'(\pi-\tilde{\delta})\,.
\end{align}
This relation with the small $\tilde{\delta}$ limit and identifying $\mathcal{E}(\varphi)= E_{J} \frac{1}{\sqrt{-h_{\varphi\varphi}}} \delta(\varphi-\pi) + \sigma$ lead to the following joint energy:
\begin{align}\label{E_J}
E_{J} =&\, 2\sigma L   \frac{\sqrt{\frac{Z_J^2}{z_+^2}-1}}{\left(1-\frac{Z_J^2}{z_-^2}\right)}\sqrt{\left(1-\frac{Z_J^2}{z_-^2}\right)+ \sigma ^2 \left(\frac{Z_J^2}{z_+^2}-1\right)}\nonumber\\
=&\,2 L \sqrt{1-\sigma ^2}\cosh ^2\left(\frac{\pi -\frac{\Delta \varphi }{2}}{z_+}\right) \sqrt{\tanh ^2\left(\frac{\pi -\frac{\Delta \varphi }{2}}{z_+}\right)-\sigma ^2} \,,
\end{align}
where $Z_J$ is the location of the joint defined by $Z_J \equiv Z(\pi)$. G. Hayward introduced a nonsmooth junction caused by a Gibbons-Hawking-like term on the junction in \cite{Hayward:1993my}. Our expression $E_J$ is the right energy expression producing the joint of the EoW brane.

It is desirable to endow a geometric meaning with this energy. A smooth configuration makes the joint energy $E_J$ vanish. So, one can expect the energy to be related to the normal vectors to the left and right parts of the brane. The following one-forms give those normal vectors near the joint:
\begin{align}
n^{(l)}= \frac{L}{Z_J}\frac{\sigma}{1-\frac{Z_J^2}{z_-^2}} \left( -dz + Z'(\pi-\tilde{\delta} ) d\varphi  \right)~,~n^{(r)}=  \frac{L}{Z_J}\frac{\sigma}{1-\frac{Z_J^2}{z_-^2}}\left( -dz + Z'(\pi+\tilde{\delta} ) d\varphi  \right)\,.
\end{align}
Note that $n^{(l)}\cdot n^{(l)} =n^{(r)}\cdot n^{(r)} = 1$. By symmetry, $Z'(\pi + \tilde{\delta})= -Z'(\pi - \tilde{\delta})$. Then, the joint energy is expressed in terms of the normal vectors as follows:
\begin{align}\label{H dual}
E_J \sqrt{g_{\bar{t}\bar{t}}}\, d\bar{t}= E_J \,e^{\bar{t}} =  L \, {}^*({n}^{(l)} \wedge {n}^{(r)} )\,,
\end{align}
where ${}^*$ denotes the 3-dimensional Hodge dual based on the orientation with $\epsilon^{\bar{t}z\varphi}=1$. So, the geometric meaning of the joint energy is the area of a parallelogram made of two normal vectors ${n}^{(l)}$ and ${n}^{(r)}$ near the joint or their outer product.

On the other hand, we again refer to \cite{Hayward:1993my} for a similar geometrical meaning. The paper shows that one may define an angle using the inner product between the normal vectors near the joint. Such a quantity is used for describing a Gibbons-Hawking-like term for a joint position. In our case, the inner product of both normal vectors is given by
\begin{align}
n^{(l)}\cdot n^{(r)} =-\left( 1 +2\sigma^2\, \frac{\left(\frac{Z_J^2}{z_+^2}-1\right)}{\left(1-\frac{Z_J^2}{z_-^2}\right)}\right)\,.
\end{align}
The negativity arises from $\mathbb{Z}_2$-symmetry about $\varphi=\pi$. Therefore, it is tempting to define an angle parameter $\gamma \equiv \cosh^{-1} ( - n^{(l)}\cdot n^{(r)} )$. The joint energy can be written as
\begin{align}
E_J = L \sinh\gamma.
\end{align}
This expression guarantees that $E_J$ is a scalar quantity. This formula is not a generic form, since the property of the normal vectors changes the definition of the angle $\gamma$.

\subsection{Double-joint EoW brane}

Now, we would like to show another possible brane configuration inside the horizon. When the embedding function $Z(\varphi)$ is a constant, {\it i.e.}, $Z(\varphi)=Z_0$, the normal vector of this EoW brane inside the horizon is a timelike vector with $\epsilon_s=-1$. Therefore, this part of the EoW brane is spacelike. Also, the junction equations become the following simple forms: 
\begin{align}
\mathcal{E}= \epsilon _d \frac{\sqrt{\frac{Z_0^2}{z_+^2}-1}}{\sqrt{1-\frac{Z_0^2}{z_-^2}}} ,~~~~\mathcal{P}= \epsilon _d  \frac{\sqrt{1-\frac{Z_0^2}{z_-^2}}}{\sqrt{\frac{Z_0^2}{z_+^2}-1}}\,\,.
\end{align}
Note that $\mathcal{P}\mathcal{E}=1$. This brane can be attached to the previous timelike-brane, like the third-column figures in Figure \ref{fig: EoW cartoon01} or Figure \ref{fig: EoW cartoon02}. The green straight lines and circles denote spacelike EoW branes, while the red curves are the timelike EoW branes in the figures. Two joints indicated by the blue dots connect these timelike and spacelike segments.

Using the result above, we can identify the spacelike brane with two joints that connect the timelike branes. From the lesson of the previous section, one can notice that the first term of (\ref{Junct 02}), including the energy density $\mathcal{E}(\varphi)$, is crucial to determine the physical properties of the joints. As one can see, this first term contains the orientation sign, $\epsilon_d$, and $|\mathcal{P}(\varphi)|$. To avoid a subtlety in the integration near a joint and to consider the simplest configuration, we set $\epsilon_d=1$ and $\mathcal{P}=\sigma$ for the spacelike brane. Adopting this choice, the location of the spacelike brane is determined by
\begin{align}\label{Z0}
Z_0 = \frac{\sqrt{1+\sigma^2}}{\sqrt{\frac{z_+^2}{z_-^2}+\sigma ^2}}\,z_+\,.
\end{align}
It is easy to show that $z_{-}\geq Z_0\geq z_{+}$. Therefore, the energy momentum tensor of the spacelike EoW brane segment is given by $\mathcal{E}=1/\sigma$ and $\mathcal{P}=\sigma$. Notably, the energy-momentum tensor of the timelike segment was introduced as $\mathcal{E}=-\mathcal{P}=\sigma$. On the other hand, the spatial component of the induced metric $h_{\varphi\varphi}$ has the opposite signs for the spacelike($h_{\varphi\varphi}>0$) and timelike($h_{\varphi\varphi}<0$) brane segments. So, we allow this change ($\sigma\leftrightarrow -\sigma $) in the pressure sign over the EoW brane configuration. However, the spatial component of the junction energy-momentum tensor $T^{(m)}_{\varphi\varphi}$ (\ref{Junct 00}) is continuous along the EoW brane. See a cartoon in Figure \ref{fig: EoW cartoon02} to see the detailed situation.

With this choice, one may consider an integration over a tiny region again near a joint as follows:
\begin{align}
\int_{\varphi_L -\tilde{\delta}}^{\varphi_L +\tilde{\delta}} d\varphi Z''(\varphi) = -Z'(\varphi_L -\tilde{\delta}),
\end{align}
where $\varphi_L$ is the left joint location ($\varphi=\varphi_L$), and we use vanishing $Z'(\varphi_L+\tilde{\delta})$ for the spacelike brane because $Z(\varphi)=Z_0$ ranging $\varphi_L\leq\varphi\leq\varphi_R$. One may notice a factor-of-2 difference from the previous case. In addition, we define the joint energy as 
\begin{align}
\mathcal{E}(\varphi)\sim  \frac{E_L}{\sqrt{|h_{\varphi\varphi}|}} \delta({\varphi-\varphi_L}) + \frac{E_R}{\sqrt{|h_{\varphi\varphi}|}} \delta({\varphi-\varphi_R})\,,
\end{align}
where `$\sim$' means the singular contributions, and $\varphi_L$ and $\varphi_R$ are the positions of the left and right joints in Figure \ref{fig: EoW cartoon02}. Then, the form of the joint energy is nothing but half of (\ref{E_J}), and its value is determined by
\begin{align}\label{EL-ER}
E_{L} = E_{R} = \sigma L   \frac{ \sqrt{\frac{Z_0^2}{z_+^2}-1} }{\left(1-\frac{Z_0^2}{z_-^2}\right)}\sqrt{\left(1-\frac{Z_0^2}{z_-^2}\right)+ \sigma ^2 \left(\frac{Z_0^2}{z_+^2}-1\right)}=\sqrt{2}L\,.
\end{align}
Here, it is interesting to note that the joint energy is independent of the tension, balancing between the brane position $Z_0$ and the tension.

To see a geometrical meaning of these joint energies, we define the normal vector for the spacelike-brane segment as follows:
\begin{align}
n^{(c)} = -\frac{L}{Z_0 \sqrt{1-\frac{Z_0^2}{z_-^2}} \sqrt{\frac{Z_0^2}{z_+^2}-1}} dz\,.
\end{align}
The normal vectors to the left and right timelike segments are given by
\begin{align}
&n^{(l)}= \frac{L}{Z_0}\frac{\sigma}{1-\frac{Z_0^2}{z_-^2}} \left( -dz + Z'(\varphi_L-\tilde{\delta} ) d\varphi  \right)\,,\nonumber\\
&n^{(r)}=  \frac{L}{Z_0}\frac{\sigma}{1-\frac{Z_0^2}{z_-^2}}\left( -dz + Z'(\varphi_R+\tilde{\delta} ) d\varphi  \right)\,.
\end{align} 
Then, the energy $E_L$ or $E_R$ is again related to the area of the parallelogram (the outer product) made of $n^{(l)}$ and $n^{(c)}$ or $n^{(c)}$ and $n^{(r)}$. One can show that the previous expression (\ref{H dual}) is allowed again as follows:
\begin{align}
&L{}^* n^{(l)}\wedge n^{(c)}= E_L \sqrt{g_{\bar{t}\bar{t}}}\, d\bar{t}=E_L e^{\bar{t}}\,,\nonumber\\
&L{}^* n^{(c)}\wedge n^{(l)}= E_R \sqrt{g_{\bar{t}\bar{t}}}\, d\bar{t}=E_R e^{\bar{t}}\,,
\end{align}
where $E_L=E_R=\sqrt{2}L$. Note that the above quantities are all evaluated at $z=Z_0$, and this location has inner product values among normal vectors as $n^{(l)}\cdot n^{(c)}=   n^{(c)}\cdot n^{(r)}=-1$.


\begin{figure}
        \centering
        \begin{subfigure}{}
            \centering
            \includegraphics[width=72mm]{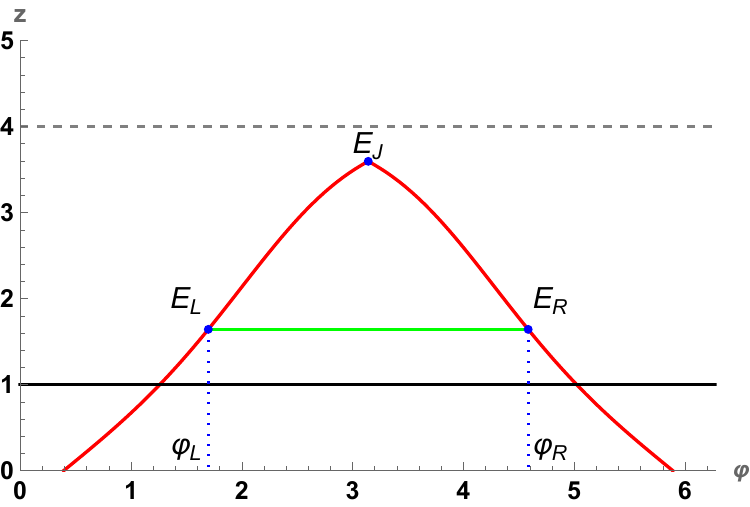}
        \end{subfigure}
        \begin{subfigure}{}
            \centering
            \includegraphics[width=65mm]{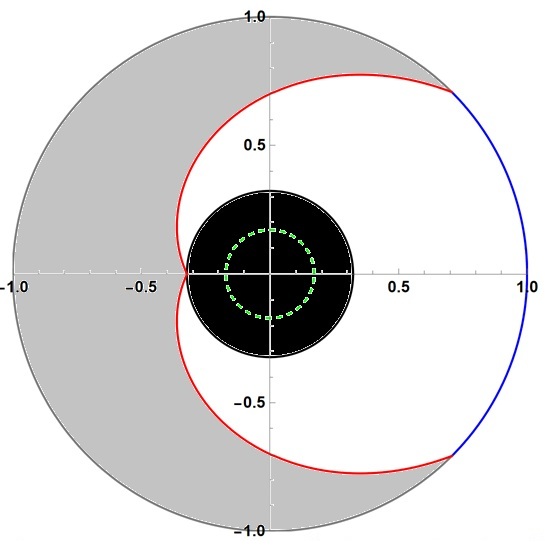}
        \end{subfigure}        
                 \caption{{\bf Overlapped two interior configurations(Left)}: We set $z_+=1$ and $z_-=4$. The red curve and the blue dot $E_J$ describe a single-joint configuration. On the other hand, a double-joint configuration consists of two lower blue dots ($E_L$ and $E_R$), the green segment, and two red segments from  $\varphi_L$ and $\varphi_R$ to the boundary ($z=0$). \\
{\bf EoW brane pinched by the horizon(Right)}: The gray region denotes the truncated spacetime. The red and blue curves are the EoW brane and the BCFT bath with volume $\mathcal{V}=L\Delta\phi$. The black circular region and the green dashed circle denote the interior of the horizon and the inner horizon as well.}\label{fig: compare cartoon00}
\end{figure}{}

\begin{figure}
        \centering
        \begin{subfigure}{}
            \centering
            \includegraphics[width=70mm]{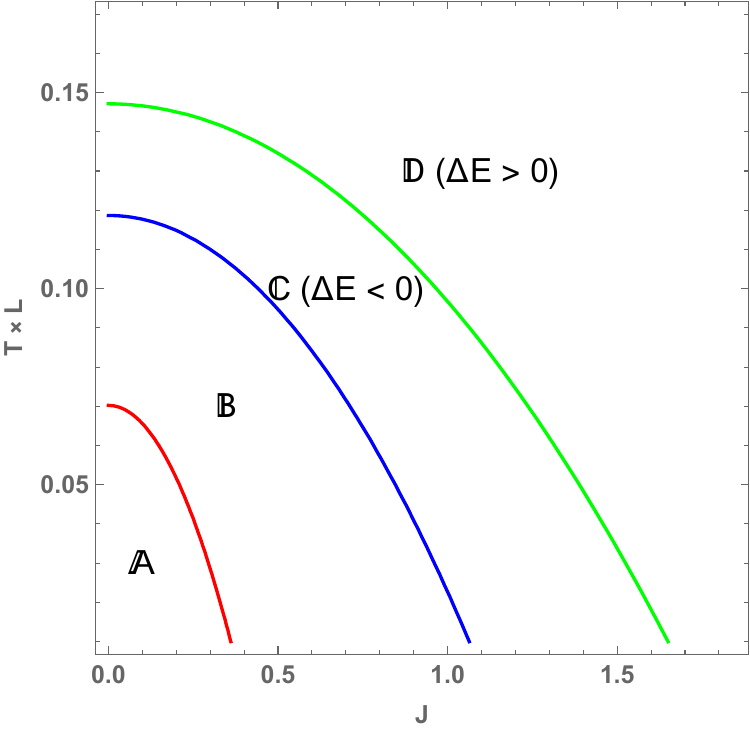}
        \end{subfigure}
                 \caption{{\bf Significant changes of the EoW branes inside the black hole:} We choose $\Delta\phi=\pi$ and $\sigma=0.6$. $\mathbb{A}$ denotes the region where the EoW brane comes out completely outside of the horizon. $\mathbb{B}$ is the region that allows only the single-joint EoW brane. In contrast, both single and double-joint configurations can exist in $\mathbb{C}$ and $\mathbb{D}$. The energy of the single-joint configuration is smaller than that of the double-joint one in the $\mathbb{C}$ region, but it is larger in $\mathbb{D}$. }\label{fig: Phase diagram}
\end{figure}{}

\subsection{Interior difference between single and double-joint EoW branes}

From the view of observers outside the horizon, it is impossible to distinguish disparate interior structures. However, in the previous sections, we found that the interior brane structure may differ even for the same configuration outside the horizon. Therefore, we provide comments on the differences between the single and double-joint configurations. Mainly, we focus on the energies of both EoW branes.
Note that our main purpose here is to demonstrate the possibility of various configurations appearing in the thermodynamic phase diagram. We therefore restrict our analysis to configurations in which
$\mathcal{P} =\sigma$
is continuous across the exterior and interior regions.

As one can see in the left figure of Figure \ref{fig: compare cartoon00}, the apex location $Z_J$, whose the joint energy $E_J$ in a single-joint configuration should be higher than the spacelike EoW brane location $Z_0$ given in (\ref{Z0}) to have both interior configurations with an identical exterior configuration. Thus, we are interested in a parameter region, where $Z_J\geq Z_0$,
\begin{align}
\frac{\sqrt{1-\sigma ^2} \tanh \left(\frac{\pi -\frac{\Delta \varphi }{2}}{z_+}\right)}{\sqrt{\sigma ^2 \left(1-\frac{z_+^2}{z_-^2}\right)+\left(\frac{z_+^2}{z_-^2}-\sigma ^2\right) \tanh ^2\left(\frac{\pi -\frac{\Delta \varphi }{2}}{z_+}\right)}} \geq  \frac{\sqrt{1+\sigma^2}}{\sqrt{\frac{z_+^2}{z_-^2}+\sigma ^2}}\,.
\end{align}
Note that $\sigma ^2 \left(1-\frac{z_+^2}{z_-^2}\right)+\left(\frac{z_+^2}{z_-^2}-\sigma ^2\right) \tanh ^2\left(\frac{\pi -\frac{\Delta \varphi }{2}}{z_+}\right)\geq 0$ is satisfied trivially. Therefore, the condition for the existence of the double-joint configuration is given by
\begin{align}
2\pi - \Delta\phi \geq 2 z_+  \tanh^{-1}\sqrt{ \frac{1+\sigma^2}{2}  }\,.
\end{align}
In addition, the apex joint of the single-joint EoW brane exists inside the horizon when the parameter region satisfies
\begin{align}
2\pi - \Delta\phi \geq 2 z_+  \tanh^{-1}\sigma\,.
\end{align}

Now, let us consider the energy difference between the single and double-joint EoW branes. We consider only the energy carried on the EoW branes. To find the difference, we define $\Delta \mathbf{E}=\mathcal{E}_{1-Joint}-\mathcal{E}_{2-Joint}$, where each energy is defined by
\begin{align}
&\mathcal{E}_{1-Joint} = \int_{\varphi_L}^{\varphi_R} d\varphi \left(\sqrt{|h_{\varphi\varphi}|} \,\sigma\right)_{z=Z(\varphi)} + E_J\,,\nonumber\\
&\mathcal{E}_{2-Joint} =\int_{\varphi_L}^{\varphi_R} d\varphi \left(\sqrt{|h_{\varphi\varphi}|} \,\frac{1}{\sigma}\right)_{z=Z_0} + E_L + E_R\,.
\end{align}
We excluded the energy contribution from the overlap region. See the left figure of Figure \ref{fig: compare cartoon00}. Here, $\varphi_L=z_+ \tanh^{-1}\sqrt{\frac{1 +\sigma^2}{2}}+\frac{\Delta\phi}{2}$ and $\varphi_R = 2\pi-\varphi_L$. The double-joint energy is given by
\begin{align}
\mathcal{E}_{2-joint} = 2 \sqrt{2} L + \frac{\sigma L}{\sqrt{1+\sigma^2} }\left(  \frac{\left(2 \pi-\Delta \varphi\right)}{z_+} + 2  \tanh ^{-1}\sqrt{\frac{1+\sigma^2}{2} }\right)\,.
\end{align}
In addition, the single-joint energy takes the following form:
\begin{align}
\mathcal{E}_{1-Joint} &=E_J +2\int_{Z_0}^{Z_J} dz \left(\frac{1}{Z'}  \sqrt{|h_{\varphi\varphi}|} \,\sigma\right)_{\varphi=\varphi(z)} \nonumber \\
&=E_J+ 2\,\sigma\int_{Z_0}^{Z_J} dz \left( \frac{L}{z} \sqrt{\frac{1}{\left(1-\frac{z^2}{z_-^2}\right)\left(\frac{z^2}{z_+^2}-1 \right)} - \frac{1}{Z'^2}\frac{1- \frac{z^2}{z_-^2}}{1-\frac{z_+^2}{z_-^2}} }\,\, \right)_{\varphi=\varphi(z)} \nonumber\\
&= 2 L \sqrt{1-\sigma ^2} \cosh ^2\left(\frac{\pi -\frac{\Delta \varphi }{2}}{z_+}\right) \sqrt{\tanh ^2\left(\frac{\pi -\frac{\Delta \varphi }{2}}{z_+}\right)-\sigma ^2}\nonumber\\
&~~+ \frac{2\sigma L}{\sqrt{1-\sigma^2}}\left\{\tan^{-1}\left( \frac{1}{\sigma} \sqrt{\tanh ^2\left(\frac{\pi -\frac{\Delta \varphi }{2}}{z_+}\right)-\sigma^2} \right)-\tan^{-1}\left(\frac{\sqrt{1-\sigma^2}}{\sqrt{2} \sigma }\right) \right\}.
\end{align}
Notably, the energies do not depend on $ z_{-}$, so this can readily be applied to the non-rotating BTZ black hole.

From the perspective of an observer residing at $z=0$ on the EoW brane, the trajectory from region $\mathbb{D}$ to $\mathbb{A}$ in Figure \ref{fig: Phase diagram}---driven by the reduction of temperature or angular momentum---unveils a profound evolution in the internal geometry of the black hole. While the observer perceives a consistent JT black hole background throughout regions $\mathbb{D}$, $\mathbb{C}$, and $\mathbb{B}$, a significant structural transformation may occur hidden behind the horizon. This change is governed by the energy competition between the single-joint and double-joint EoW brane configurations; specifically, while the double-joint configuration is energetically favored in the high-temperature $\mathbb{D}$ region ($\Delta\mathbf{E} > 0$), the single-joint configuration becomes dominant in region $\mathbb{C}$ ($\Delta\mathbf{E} < 0$). This internal transition differs inherently from the boundary-affected Hawking-Page transitions discussed in \cite{Takayanagi:2011zk, Fujita:2011fp}, as it transpires entirely within the interior. Although an asymptotic observer may not detect a sharp phase-transition signal, these dynamics suggest that the EoW brane serves as a highly sensitive probe for deciphering the hidden microstate configurations and nontrivial structural transitions within holographic black holes.

In addition to this change, observers at $z=0$ in the opposite positions on the EoW brane can finally communicate with each other in the low-temperature and small-angular-momentum region $\mathbb{A}$. Therefore, the EoW geometry changes from two black holes to one connected spacetime with two $AdS_2$ boundaries at the borderline between $\mathbb{A}$ and $\mathbb{B}$ regions. This sudden out-of-horizon exposure of the EoW brane for the non-rotating BTZ was discussed in our previous work \cite{Kim:2023adq}. This exposure event happens when the following condition is satisfied:
\begin{align}\label{pinch equation}
\tanh^{-1}\sigma =\frac{1}{2z_+} \left(2\pi-\Delta\phi\right)\,.
\end{align}
This implies that the boundary entropy, equivalently, the shadow entropy reaches its maximum value. We depict an EoW brane of this case in the right figure of Figure \ref{fig: compare cartoon00}. Interestingly, this occurs at a finite temperature.

Intriguingly, the joint energy (\ref{E_J}) of this particular configuration vanishes when the joint touches the horizon,  $Z_J=z_+$. The configuration shows a pinched single brane by the horizon. Thus, the disappearing joint energy may transfer to the singularly pinched shape. To see what happens to this pinching point, one can calculate the difference of the normal vectors approaching the joint from both directions. The difference turns out to be
\begin{align}
\left(n^{(l)}-n^{(r)}\right)^2_{z=z_+}= 4\,.
\end{align}
This does not vanish. This observation confirms that there is an excessive angle at the joint, so a kind of conical singularity lies at the pinching point. Therefore, it can be seen that joint energy was transferred to the induced geometric energy of the EoW brane at the pinching point.

\section{Conclusion}

In this paper, we investigate the holographic properties of dynamical EoW branes within the background of a rotating BTZ black hole. By solving the junction equations, we demonstrate that the induced geometry on these branes precisely maps to an effective two-dimensional JT gravity system. This mapping allows us to rigorously derive the first law of thermodynamics for the associated BCFT, incorporating the boundary degrees of freedom as dynamical variables. Building upon our previous investigation into non-rotating systems \cite{Kim:2023adq}, the present work successfully incorporates angular momentum to establish a more general holographic dictionary. The primary achievement of this extension is encapsulated in Equations (\ref{U PV}) and (\ref{first law final}), which formally bridge the rotating BTZ thermodynamics with geometric cutoffs to the BCFT. These two expressions of the first law are based on (\ref{first VdP}) and (\ref{first law JT SYK}), respectively. Furthermore, the appropriate treatment of the EoW brane contribution developed in our framework could potentially provide a natural way to formulate the consistent first law of thermodynamics for other brane-involved setups, such as accelerating BTZ black holes \cite{Arenas-Henriquez:2022www, Arenas-Henriquez:2023hur}.

Through the framework of black hole chemistry, we identify the tension-dependent thermodynamic pressure and volume terms in the JT black hole. Using these parameters, the first law of JT black hole (\ref{first VdP}) can be constructed. On the other hand, we establish a connection between the macroscopic thermodynamics of the JT system and the microscopic parameters of the SYK model. By interpreting the variation of brane tension as a change in the average coupling strength $\mathcal{J}_{\text{SYK}}$, we formulate (\ref{first law JT SYK}). These variation laws hold good for the induced geometry of an individual EoW brane. Then, to incorporate these expressions into the BCFT bulk first law (\ref{first law without boundary}), one needs to redefine the conjugate thermodynamic quantities by using the wrapping factor (\ref{wrapping factor}). Then, one can arrive at (\ref{U PV}) and (\ref{first law final}).

We also examine the equivalence between the boundary and shadow entropies by analyzing HRT surfaces in the rotating BTZ black hole. Our analysis verifies that the shadow entropy, defined by the horizon area obscured by the EoW branes, is exactly equivalent to the boundary entropy based on a particular spacelike sheet. This equivalence remains robust even in the presence of angular momentum, highlighting the universality of BCFT boundary entropy. It provides strong evidence that the geometric profile of the EoW brane accurately captures the microscopic degrees of freedom at the boundary.

Finally, we explore the black hole's interior structure by extending the EoW branes beyond the event horizon. We identify two representative configurations: the single-joint and double-joint structures. By comparing their respective energies, we propose an internal transition that may occur within the horizon, triggered by variations in temperature and angular momentum. These findings imply that the black hole interior may host fundamental structural transitions inaccessible to asymptotic observers.

As a final comment, one of the most crucial implications in the present work is a hidden link between the SYK model and the Cardy state \cite{Cardy:1989ir}, which governs the boundary condition of BCFT. This mysterious connection may help unveil the origin of gravity in a quite compact theoretical model. We leave it as a future study.

\section*{Appendix}

\subsection*{A. Boundary quantities}

The renormalized Lorentzian action is given by
\begin{align}
S =\frac{1}{16\pi G} \int d^3 x \sqrt{-g} \left( R + \frac{2}{L^2} \right) + \frac{2}{16\pi G}\int_{r=\Lambda_c} d^2 x\sqrt{-\hat{h}}\left(K - \frac{1}{L} \right)\,, 
\end{align}
where $\hat{h}=\det{\hat{h}_{ab}}$, and $\hat{h}_{ab}$ is the induced metric with the following ADM decomposition along the holographic direction:
\begin{align}
ds^2 = \mathcal{N}^2 dr^2 + \hat{h}_{ab}\left(dx^a + \mathcal{N}^a dr \right)\left(dx^b + \mathcal{N}^b dr \right)\,.
\end{align} 
Also, $K$ is the extrinsic curvature of the holographic screen $r=\Lambda_c$. The holographic energy-momentum tensor for the boundary system is
\begin{align}
\left< T_{ab} \right> =\frac{1}{16\pi G L} \left(
\begin{array}{cc}
 M  &   - JL  \\
 -JL  & M L^2 \\
\end{array}
\right)\,.
\end{align}
The boundary system has the Minkowski metric, $ds^2 = - dt^2 + L^2 d\phi^2$. Thus, the trace of the energy-momentum tensor vanishes. The on-shell action obtained by the BTZ solution is 
\begin{align}
S^{\text{on-shell}} = \frac{1}{16\pi G L} \int dt \int d \phi L \left( \frac{2 r_+^2}{L^2} - M \right)\,.  
\end{align} 
Applying analytic continuation, the free energy $\mathcal{W}$ is related to the on-shell action by $-S^{\text{on-shell}}=\mathcal{W}/T$. Thus,
\begin{align}
\frac{\mathcal{W}}{T} = -\int_0^{\frac{1}{T}} d\tau_E  \int_0^{\Delta\phi}  d\phi\,L \, \bar{w} = -\frac{\mathcal{V}}{T}\frac{(r_+^2-r_-^2)}{16 \pi  G L^3}\,,
\end{align}
where $\bar{w}$ is the free energy density. For a system with translational invariance, the pressure of the boundary system is given by $\mathcal{P}=-\bar{w}$. Therefore, the pressure can be written as
\begin{align}
\mathcal{P} = \frac{1}{16\pi G L} \left( \frac{1}{z_+^2} - \frac{1}{z_-^2} \right)\,.
\end{align}
Note $\mathcal{P}\neq T_{\phi\phi}$ due to the Casimir pressure contribution. $z$ and $r$ are related with $z=L/r$.

\section*{B. Validity of the Junction Equation}

The junction equation governs the EoW brane. It is nothing but the Einstein equation in the limit of a thin energy-momentum distribution. Therefore, such a configuration backreacts on the background geometry in principle. However, the thin distribution means we deal with the EoW brane as a probe. To remain in the probe limit, the tension $\sigma/L$ should not be large, as the referee comments (we took $\sigma$ as a dimensionless positive parameter to avoid negative energy). The unit of tension is $L^{-1}$, and the curvature radius L is large in the gauge/gravity duality. Therefore, the tension must be a finite number $\sigma$ times $L^{-1}$. Also, we consider EoW branes attached to the boundary of $AdS_3$ space. This tension is associated with an angle between the EoW brane and the AdS boundary. The allowed parameter region for this probe configuration is $\sigma \leq 1$.

~~We also restrict our study to the configuration where the induced metric is that of the JT black hole. In this configuration, the two EoW branes touch the horizon.  The pinched configuration in the right figure of Figure 6 gives the boundary of the tension range under consideration. If the two branes have different tensions, the range for the pinched case is more complicated. However, the case of equal tension gives rise to a simple range using (\ref{pinch equation}). We can easily find that the tension range for well-defined thermodynamics is as follows:
\begin{align*}
0<\sigma < \sigma_p=\tanh \left(\frac{2\pi-\Delta\phi}{2z_+}\right)<1\,,
\end{align*}
where $\Delta\phi$ is the volume angle of the BCFT, smaller than $2\pi$. Therefore, we consider $0<\sigma<\sigma_p$ as the parameter region where the thermodynamics is defined. 

~~In particular, in the $\sigma=1$ (tension is $1/L$.) case, the EoW branes are parallel to the boundary of the AdS space. Thus, the branes can't touch the horizon, and their induced metrics don't have a horizon. This is not our interest in the present paper. Even on the JT gravity side, this case gives rise to flat JT gravity (JT gravity with the vanishing cosmological constant). This should also be dealt with for a different research purpose. When $\sigma_p<\sigma<1$, the two EoW branes make a joint outside the horizon. We study this case in a future project.

\section*{Acknowledgments}
K. Kim appreciates San-Jin Sin, Geng, and Svesko for constructive comments. 
This work is supported by the Basic Science Research Program through the National Research Foundation of Korea (NRF) grant No. NRF-2022R1A2C1010756 (Y. Seo, HE), NRF-2019R1A2C1007396 (K. Kim). YS was partially supported by NRF grant funded by the Korea government(MSIT) No. RS-2026-25588089. HE was partially supported by Basic Science Research Program through the NRF funded by the Ministry of Education (NRF-2022R1I1A1A01068833). J. H. L. was supported by the NRF of Korea grant funded by the Korean government (No. NRF-2021R1C1C2008737).
We thank APCTP, Pohang, Korea for their hospitality during the Focus Program [APCTP-2026-T04] from which this work greatly benefited. We are the members of Korea Research Network for Theoretical Physics and thank the other members for helpful discussions.


\end{document}